\documentclass[11pt,letterpaper,preprintnumbers]{article}
\usepackage{jheppub}
\usepackage{bbm}
\usepackage{epsfig,bm}
\usepackage{graphicx,comment}

\DeclareMathOperator{\tr}{tr}
\DeclareMathOperator{\Pf}{Pf}

\title{
Orbifold equivalence for finite density QCD  \\
and effective field theory
} 

\author[a]{Aleksey Cherman}
\author[b]{and Brian C.~Tiburzi}

\emailAdd{a.cherman@damtp.cam.ac.uk}
\affiliation[a]{%
Department of Applied Mathematics and Theoretical Physics,\\ 
Cambridge University, \\
Cambridge CB3 0WB,
UK
}

\emailAdd{bctiburz@mit.edu}
\affiliation[b]{%
Center for Theoretical Physics,\\
Massachusetts Institute of Technology,\\ 
Cambridge, Massachusetts 02139,
USA
}

\abstract{%
In the large 
$N_{c}$ 
limit, 
some apparently different 
gauge theories turn out to be equivalent due to large 
$N_{c}$ 
orbifold equivalence. 
We use effective field theory techniques to explore orbifold equivalence, focusing on the specific case of a recently discovered relation between an 
$SO(2N_{c})$ 
gauge theory and QCD.    
The equivalence to QCD has been argued to hold at finite baryon chemical potential, 
$\mu_{B}$, 
so long as one deforms the 
$SO(2N_{c})$ 
theory by certain ``double-trace''  terms.  
The deformed 
$SO(2N_c)$
theory can be studied without a sign problem in the chiral limit, 
in contrast to
$SU(N_{c})$ 
QCD at finite 
$\mu_{B}$.  
The purpose of the double-trace deformation in the 
$SO(2N_{c})$ 
theory is to prevent baryon number symmetry from breaking spontaneously at finite density, 
which is necessary for the equivalence to large 
$N_{c}$ 
QCD to be valid.  
The effective field theory analysis presented here clarifies the physical significance of double-trace deformations, 
and strongly supports the proposed equivalence between the deformed 
$SO(2N_{c})$ 
theory and large 
$N_{c}$ 
QCD at finite density.  
}

\begin{document}

\def\a{{\alpha}}
\def\b{{\beta}}
\def\d{{\delta}}
\def\D{{\Delta}}
\def\X{{\Xi}}
\def\e{{\varepsilon}}
\def\g{{\gamma}}
\def\G{{\Gamma}}
\def\k{{\kappa}}
\def\l{{\lambda}}
\def\L{{\Lambda}}
\def\m{{\mu}}
\def\n{{\nu}}
\def\o{{\omega}}
\def\O{{\Omega}}
\def\S{{\Sigma}}
\def\s{{\sigma}}
\def\th{{\theta}}

\def\ol#1{{\overline{#1}}}
\def\eqref#1{{(\ref{#1})}}

\def\diag{\text{diag}}

\def\cF{{\mathcal F}}
\def\cS{{\mathcal S}}
\def\cC{{\mathcal C}}
\def\cB{{\mathcal B}}
\def\cT{{\mathcal T}}
\def\cQ{{\mathcal Q}}
\def\cL{{\mathcal L}}
\def\cO{{\mathcal O}}
\def\cA{{\mathcal A}}
\def\cQ{{\mathcal Q}}
\def\cR{{\mathcal R}}
\def\cH{{\mathcal H}}
\def\cW{{\mathcal W}}
\def\cM{{\mathcal M}}
\def\cD{{\mathcal D}}
\def\cN{{\mathcal N}}
\def\cP{{\mathcal P}}
\def\cK{{\mathcal K}}
\def\cI{{\mathcal{I}}}
\def\cJ{{\mathcal{J}}}

\def\be{\begin{equation}}
\def\ee{\end{equation}}

\maketitle%

\section{Introduction}								             %

Our understanding of four-dimensional non-Abelian gauge theories such as QCD is limited by the fact that they are strongly coupled at low energies, making analytic insights difficult to obtain.  For three-color QCD, the only known approach to computing generic observables beyond perturbation theory is based on lattice Monte Carlo simulations.  In the limit of a large number of colors, 
$N_{c} \to \infty$, 
however, QCD dramatically simplifies~\cite{tHooft:1973jz,Witten:1979kh}. At large $N_{c}$, QCD becomes a theory of stable, weakly-interacting mesons and glueballs, with baryons emerging as solitons of meson fields.  Remarkably, the large $N_{c}$ world does not appear to be too different from our $N_{c}=3$ world for many observables, so that one can hope to use a $1/N_{c}$ expansion to make predictions for real-world QCD.  Unfortunately, despite the simplifications occurring at large $N_{c}$ in QCD, it is not known how to solve the $N_{c} =\infty$ theory, and the predictions of $1/N_{c}$ expansions for QCD are mostly of a qualitative nature.\footnote{There are a few special cases in which one can do better and make semi-quantitative predictions using $1/N_{c}$ expansions. Such predictions tend to show good agreement with experimental data~\cite{Dashen:1993as,Dashen:1993jt,Jenkins:1995td,Cherman:2009fh,Lebed:2010um}.}

Still, there is reason to be optimistic that large $N_{c}$ expansions can be more than just qualitatively useful.  
Large $N_{c}$ gauge theories often have remarkable properties.  
In some cases, 
which unfortunately are not known to include QCD, 
one \emph{can} solve the $N_{c}=\infty$ theory by using strong-weak dualities such as AdS/CFT~\cite{Maldacena:1997re,Witten:1998qj,Gubser:1998bc}, 
which relate some strongly-coupled, 
four-dimensional gauge theories at large $N_{c}$ to tractable weakly-coupled 
string theories living in ten dimensions.  A different notion (but one historically connected~\cite{Kachru:1998ys} to AdS/CFT) is that of strong-strong dualities,
which usually go by the name of large $N_{c}$ orbifold equivalences~\cite{Bershadsky:1998cb,Schmaltz:1998bg,Strassler:2001fs,Armoni:2003gp,Kovtun:2004bz,Kovtun:2003hr}.  Orbifold equivalences, which will be the focus of this paper, connect gauge theories with different gauge groups and matter content.  Large $N_{c}$ orbifold-equivalent theories have a set of correlation functions that coincide at large $N_{c}$;  the associated observables are referred to as neutral.   Hence if an observable of interest in one theory is in the neutral sector of an orbifold equivalence, one can in principle use a large-$N_{c}$ equivalent theory to compute it.  If the calculation of such an observable is easier in one of the theories, such an equivalence would be a very useful tool.     
Unfortunately, when one of the theories involved in an orbifold equivalence is strongly coupled, the others are as well.  
Thus large 
$N_{c}$ 
orbifold equivalences are less obviously useful than strong-weak dualities such as AdS/CFT.

The advantage of large $N_{c}$ orbifold equivalences is that they can apply directly to the large $N_{c}$ versions of a gauge theory which is known to be important for understanding the real world, namely QCD.  For instance, it has been shown that $N_{f}=1$ QCD is large-$N_{c}$ equivalent to $\mathcal{N}=1$ super-Yang-Mills theory, if one extrapolates away from $N_{c}=3$ with quarks transforming in the two-index antisymmetric representation of color, instead of the usual choice of the fundamental representation~\cite{Armoni:2003gp,Armoni:2004uu}.  (The two representations coincide at $N_{c}=3$, but differ at large $N_{c}$.)  Using the powerful tools available for supersymmetric theories, this ``orientifold equivalence'' allows one to make quantitative predictions for a large $N_{c}$ limit of QCD~\cite{Armoni:2003gp,Armoni:2004uu}.

In addition to using orbifold equivalences to obtain direct analytic insights, one can also hope to use equivalences in numerical studies of QCD.  Probably the best-known idea in this direction is that of Eguchi-Kawai reduction and large $N_{c}$ volume independence~\cite{Eguchi:1982nm,Bhanot:1982sh,GonzalezArroyo:1982hz}, which can be thought of in terms of large $N_{c}$ orbifold equivalences~\cite{Kovtun:2007py}.  Large $N_{c}$ volume independence relates gauge theories in different physical volumes provided they remain in their center-symmetric phases.  Aside from enabling some analytic insights into the behavior of gauge theories~\cite{Simic:2010sv}, large $N_c$ volume independence allows lattice calculations in small volumes to access information about infinite-volume physics~
\cite{Narayanan:2003fc}.    
%

Another recent suggestion, 
which will be our primary focus here, 
is to use large $N_{c}$ orbifold equivalence to enable lattice Monte Carlo studies of QCD at finite baryon number density~\cite{Cherman:2010jj}.  
It is notoriously difficult to get any insight into the behavior of QCD at finite density from first principles.  
Away from asymptotically large densities, where the asymptotic freedom of QCD enables reliable perturbative calculations~
\cite{Rajagopal:2000wf,Alford:2007xm},  
%
finite-density QCD is strongly coupled.  Unfortunately, in contrast to the situation at zero density, lattice Monte Carlo methods are not available once one turns on a chemical potential for quark number,
$\mu$\footnote{The baryon number chemical potential $\mu_{B}$ is related to the quark number chemical potential,  
$\mu$,  
by a factor of 
$N_c$, 
namely 
$\mu_B = N_c \mu$. 
We will use the quark number chemical potential throughout, 
but will sometimes be sloppy and refer to it as the baryon chemical potential, since in QCD once $\mu$ exceeds a critical value of order $\Lambda_{QCD}$ one expects to produce nuclear matter, which consists of baryons.}.   
At finite 
$\mu$, 
the fermion determinant  in QCD becomes complex, 
and importance sampling can no longer be utilized.  
This is known as the fermion sign problem. The idea of Ref.~\cite{Cherman:2010jj} is to use large $N_{c}$ orbifold equivalence to dodge the sign problem by working with a theory which is large-$N_{c}$-equivalent to QCD, but does not have a sign problem at finite density.  One could then hope to use lattice Monte Carlo simulations of the orbifold-equivalent theory to learn about the properties of large $N_{c}$ QCD at finite density.   At the least, such studies would be of great theoretical interest.   The phenomenological usefulness  of learning about the behavior of large $N_{c}$ QCD depends on the closeness of the large $N_{c}$ world to our $N_{c}=3$ one, which is a subtle question discussed in e.g. Refs.~\cite{Shuster:1999tn,Park:1999bz,Frandsen:2005mb,Buchoff:2009za,McLerran:2007qj,Torrieri:2010gz}.

To be specific, 
the claim of Ref.~\cite{Cherman:2010jj} is that 
$SU(N_{c})$ 
gauge theory with 
$N_{f}$ 
flavors of fundamental-representation Dirac fermions, which is simply QCD when 
$N_{c}=3$, 
is orbifold-equivalent to 
$SO(2N_{c})$ 
gauge theory with 
$N_{f}$ 
vector-representation Dirac fermions in the large 
$N_{c}$ 
limit.  
The 
$SO(2N_{c})$ 
theory does not have a sign problem at finite 
$\mu$.  
A key subtlety, 
however,  
is that the equivalence of the 
$SO(2N_{c})$ 
theory to large 
$N_{c}$ 
QCD only holds as long as the 
$U(1)_{B}$ 
symmetry of the 
$SO(2N_{c})$
theory is not spontaneously broken.  
Unfortunately, 
the 
$U(1)_{B}$ 
symmetry of the 
$SO(2N_{c})$ 
theory \emph{does} break when the chemical potential exceeds half the pion mass,
$\mu \ge m_{\pi}/2$, 
as was suggested in~\cite{Cherman:2010jj} 
on general grounds, 
and is explicitly shown here in Sec.~\ref{sec:LowEnergyAnalysis}.

The issue of symmetry-breaking phase transitions invalidating large 
$N_{c}$ 
orbifold equivalences is a notoriously common difficulty 
especially when non-supersymmetric theories are involved (see e.g.~\cite{Tong:2002vp,Armoni:2005wta,Kovtun:2005kh}).  
For instance, 
what often spoils large 
$N_{c}$ 
volume independence is the breaking of center symmetry.  
In many QCD-like theories,  
center symmetry breaks in small volumes, 
invalidating the orbifold equivalences connecting large and small volume theories.  
In this context, 
\"Unsal and Yaffe~\cite{Unsal:2008ch} 
proposed a very clever way to protect center symmetry and rescue large 
$N_{c}$ 
volume independence.    
Their prescription was to deform the small-volume theory by adding certain double-trace terms to the action, 
which prevent the center-symmetry-breaking phase transition at small volumes.  This protects the large 
$N_{c}$ 
equivalence between large-volume and deformed small-volume theories for arbitrarily small volumes.  
In the current context, 
where it is a 
$U(1)_{B}$ 
symmetry that breaks, 
Ref.~\cite{Cherman:2010jj} 
proposed generalizing the idea of~\cite{Unsal:2008ch} 
to protect the 
$U(1)_{B}$ 
symmetry by including the analogue of double-trace deformations, which in this case take the form of certain four-quark operators. 
It was argued that the deformed version of the 
$SO(2N_{c})$
theory should maintain its 
$U(1)_{B}$ 
symmetry past 
$\mu = m_{\pi}/2$, 
while remaining orbifold-equivalent to QCD at large 
$N_{c}$.
All of this occurs without reintroducing a sign problem in the chiral limit.%

The validity of the equivalence proposed in~\cite{Cherman:2010jj} 
would thus provide a potential tool to study QCD at finite baryon density in the large 
$N_c$ 
limit. 
The specific deformation suggested to protect $U(1)_{B}$ 
without reintroducing a sign problem makes the tree-level action non-positive-definite, 
and one may justifiably wonder whether it indeed does the job it needs to do.  
For completely general values of 
$\mu$,
the validity of the equivalence can only be demonstrated by studying the deformed
$SO(2N_{c})$ 
theory non-perturbatively with lattice Monte Carlo techniques. 
However, 
as long as 
$\mu \sim m_\pi$ 
and the quark masses are small compared to the strong interaction scale,
$\Lambda_{SO(2N_{c})}$, 
the low-energy dynamics of the 
$SO(2N_{c})$ 
theory can be studied systematically using a low-energy effective field theory (EFT) analysis. 
Working in the EFT, 
one can determine the effects of the deformation
\emph{non-perturbatively} in the 't Hooft coupling 
$\lambda =  g_{\text{YM}}^{2} \, N_{c} $, 
where 
$g_{\text{YM}}$
is the Yang-Mills coupling.  
The analysis of the effects of the deformation on the vacuum structure of the theory is taken up in Sec.~\ref{sec:VacuumOrientation}.  
The EFT analysis allows us to develop a simple physical picture of the effects of $U(1)_{B}$-preserving deformations: 
they simply raise the masses of the particles that would otherwise condense and break 
$U(1)_{B}$.  
In the deformed theory, 
we find that the chemical potential at which the 
$U(1)_{B}$ 
symmetry breaks can be increased as far as one likes past the value
$\mu =m_{\pi}/2$, 
at least as long as one remains within the range of validity of the low-energy EFT.

The paper is organized as follows.  
We begin by briefly reviewing the orbifold projection that connects the 
$SO(2N_{c})$ 
theory to QCD in Sec.~\ref{sec:OrbifoldReview}, 
and discuss two particular deformations that are the focus of this work. 
We then sketch in Sec.~\ref{sec:HadronsAndOrbifolds} a simple way to understand the conditions necessary for orbifold equivalences to hold.  
Next, 
we construct the low-energy effective theory for the deformed 
$SO(2N_{c})$ 
theory in Sec.~\ref{sec:LowEnergyAnalysis}.  
Two particular deformations are considered: 
a chirally symmetric deformation, 
and a non-symmetric deformation. 
Spontaneous symmetry breaking in the presence of such deformations is taken up 
in Sec.~\ref{sec:VacuumOrientation}, 
where the orientation of the vacuum is determined. 
We summarize our findings by comparing the low-energy properties of the deformed
$SO(2N_{c})$
theory and large 
$N_{c}$ 
QCD in Sec.~\ref{sec:ComparisonToQCD}.  
Our results are consistent with the predictions of~\cite{Cherman:2010jj}, 
namely they support the validity of the equivalence at the non-perturbative level.  
We conclude by outlining some possible directions for future research in Sec.~\ref{sec:ComparisonToQCD}.
Some technical details are relegated to the Appendices. 
In particular, 
we present a conjecture regarding the preservation of certain vectorial symmetries of the deformed
$SO(2N_{c})$ 
theories in Appendix~\ref{AppendixA}, 
and provide an argument as to why this conjecture is plausible. 
(Our analysis in the main text does not rely on this argument, but the results nevertheless support the conjecture.) 
Some of the technical details related to determining the vacuum alignment
in the non-symmetrically deformed theory are collected in Appendix~\ref{AppendixB}.

\section{From $SO(2 N_c)$ gauge theory to QCD}
\label{sec:OrbifoldReview}

\subsection{Orbifolding $SO( 2 N_c)$ gauge theory}

The Euclidean-space Lagrangian of the \emph{undeformed} 
$SO(2N_{c})$ 
gauge theory is
\be 
\label{eq:SOLagrangian}
\mathcal{L}_{SO} 
=
\frac{1}{4 g_{SO}^{2} } \tr \, F_{\mu \nu}^2
+ 
\sum_{a =1}^{N_{f}}
\bar{\psi}_{a} (\gamma_{\mu} D_{\mu} + m +\mu\gamma_{4}) \psi_{a} ,
\ee
where 
$F_{\mu \nu}$ 
is the field strength with 
$\mu = 1,\ldots,4$ 
denoting the Euclidean space index, 
$D_{\mu} = \partial_{\mu}+iA_{\mu}$ is the covariant derivative in the fundamental representation with 
$A_\mu = A_\mu^i t^i$, 
and
$t^i = - (t^i)^T$ are the generators of 
$SO( 2 N_c)$. 
We take the 
$N_{f}$ 
flavors to have a common bare mass $m$, 
and 
$\mu$ 
is the quark number chemical potential.  At the hadronic level, the most striking difference of this theory from QCD is that it has meson-like particles charged under baryon number.  This can be traced to the fact that making the gauge group $SO(2N_{c})$ rather than $SU(N_{c})$ means that in addition to color singlet operators of the form $\bar{\psi} \cdots \psi$, which couple to the usual mesons which are not charged under $U(1)_{B}$, one can write down color-singlet ``diquark'' operators of the form $\psi^{T} C \cdots \psi$, which couple to particles that are charged under $U(1)_{B}$.  Following \cite{Cherman:2010jj}, we will refer to the baryon-number-charged mesons by prepending a ``b'' to the names of the QCD-like mesons they resemble: bpions, b$\rho$ mesons, and so on.

The reason the 
$SO(2N_{c})$ 
theory is large-$N_{c}$ 
equivalent to QCD can be traced to the fact that there is an orbifold projection that takes the action of the 
$SO(2N_{c})$ 
 theory to the action of QCD~\cite{Cherman:2010jj}.   The physical interpretation of orbifold projections and why they result in equivalent theories is discussed in Sec.~\ref{sec:HadronsAndOrbifolds}, and for now we simply describe the mechanics of the projection.   An orbifold projection is defined by picking some discrete subgroup of the symmetry group of the 
$SO(2N_{c})$ 
theory, and then discarding all of the degrees of freedom not invariant under the action of that discrete symmetry.  
For this application, 
we pick the discrete subgroup of the 
$SO(2N_{c}) \times U(1)_{B}$ 
symmetry generated by 
$J = i \sigma_{2} \otimes \mathbbm{1}_{N_{c}} \in SO(2N_{c})$ 
and 
$\omega = e^{i\pi/2} \in U(1)_{B}$.
The action of this group on the fields themselves results in a
$\mathbb{Z}_{2}$ 
symmetry, 
with the fields transforming as
$A_{\mu} \rightarrow J A J^{T}$
and
$\psi \rightarrow \omega J \psi$.  
Setting to zero all of the non-$\mathbb{Z}_{2}$-invariant degrees of freedom in the 
$SO(2N_{c})$ 
 theory, one obtains the Lagrangian of a gauge theory with the same number of quark flavors, but with an 
 $SU(N_{c})$ gauge group\footnote{Strictly speaking, the daughter theory has a $U(N_{c})$ gauge group, but the difference between $U(N_{c})$ and $SU(N_{c})$ is a $1/N_{c}^{2}$ correction.}, provided one identifies 
 $g_{SU}^{2} = g_{SO}^{2}$.  Note that all of the diquark operators have charge $-1$ under the projection symmetry and are annihilated during the projection. 
 The bmesons are hence not in the common sector.  
 The usual mesons, meanwhile, are all in the neutral sector because they have vanishing 
$\mathbb{Z}_{2}$ charge.

Kovtun, \"Unsal, and Yaffe~\cite{Kovtun:2003hr,Kovtun:2004bz} 
showed that in order for an orbifold projection to yield a pair of orbifold-\emph{equivalent} theories, 
the symmetries involved in the projection must not be broken spontaneously.    
We give a simple way to understand the conclusions of~\cite{Kovtun:2003hr,Kovtun:2004bz} in Sec.~\ref{sec:HadronsAndOrbifolds}, 
and the results of Sec.~\ref{sec:LowEnergyAnalysis} provide an explicit illustration of the workings of orbifold equivalence.   
It is not hard to see how orbifold equivalence can get in trouble in the context of the 
$SO(2N_{c})$ 
theory at finite 
$\mu$.  
The 
$SO(2N_{c})$ 
theory contains bmesons in its spectrum, 
and if a large enough chemical potential leads them to condense, the large-$N_{c}$ equivalence will fail.  
As we will see in Sec.~\ref{sec:LowEnergyAnalysis}, 
the lightest particle in the undeformed theory that is charged under 
$U(1)_{B}$ is the bpion, 
which has mass 
$m_{\pi} \sim \sqrt{m_{q}}$ 
at 
$\mu=0$, 
the same as a pion.  
Once the chemical potential exceeds half the mass of the pions, 
the bpions condense and break 
$U(1)_{B}$, 
invalidating the equivalence of the 
$SO(2N_{c})$ 
theory to QCD
for 
$\mu > m_{\pi}/2$.

\subsection{Deforming $SO( 2 N_c)$ gauge theory}

To avoid the problems with bpion condensation, one can consider a \emph{deformed} 
$SO(2N_{c})$ 
theory, with the Lagrangian
\begin{equation} \label{eq:L}
\cL^\prime_{SO} 
=
\cL_{SO}
+ 
V (\ol \psi, \psi),
\end{equation}
where  $V (\ol \psi, \psi)$ is the deformation potential. 
The simplest deformation one can imagine is given by~\cite{Cherman:2010jj}
\begin{equation} \label{eq:deformation}
V( \ol \psi, \psi) 
= 
\mathfrak{C}^2
\sum_{a,b=1}^{N_f}
S_{ab}^\dagger S^{\phantom{\dagger}}_{ab} 
,
\end{equation}
where $\mathfrak{C}$ is a new parameter with the dimensions of inverse mass, 
while the operator
$S_{ab}
\equiv 
\psi_a^T C \gamma_5 \psi_b$,
has the quantum numbers of a scalar  bmeson.\footnote{%
In the undeformed theory, 
the conjugacy relation expressed in Eq.~\eqref{eq:Conjugate} 
in conjunction with QCD inequalities~%
\cite{Weingarten:1983uj,Nussinov:1983vh,Witten:1983ut}
can be used to demonstrate that
scalar bmesons are the lightest particles~%
\cite{Kogut:1999iv}. 
Such a demonstration does not rely on knowledge of the spontaneous symmetry breaking pattern.  
}  
 The Euclidean charge conjugation matrix is given by  
$C = \gamma_2 \gamma_4$.   
At the classical level, 
the deformation gives a repulsive interaction for two quarks coupled in the scalar diquark channel. 
In turn, 
this should induce a mass for scalar bmesons, 
which should lead to an obstruction in the formation of a scalar bmeson condensate.
This is exactly what we want, 
because the bpions turn out to be scalars under parity.
To make this schematic argument rigorous, 
we will turn to an EFT analysis below.

In the context of the EFT, 
which is based upon the pattern of chiral symmetry breaking at low energies, 
it turns out to be convenient to consider alternate deformations. 
To this end,  
we consider two other deformations: 
a chirally symmetric deformation specified by the potential
$V_+(\ol \psi, \psi)$, 
and a chirally non-symmetric deformation specified by
$V_-(\ol \psi, \psi)$, 
with
\begin{equation} 
\label{eq:pmdeformations}
V_\pm(\ol \psi, \psi)
=
\mathfrak{C}^2
\sum_{a,b=1}^{N_f}
\left(
S_{ab}^\dagger S^{\phantom{\dagger}}_{ab} 
\pm
P_{ab}^\dagger P^{\phantom{\dagger}}_{ab}
\right)
,\end{equation}
where the operator
$P_{ab} \equiv \psi_a^T C \psi_b$
has the quantum numbers of a pseudo-scalar bmeson. 
The chirally symmetric deformation should penalize both scalar and pseudo-scalar bmesons from condensing. 
It is less clear what happens for the non-symmetric deformation, 
as there is a competition between the two terms.  
The EFT analysis can be used for both cases, 
a linear combination of which is the simple deformation in Eq.~\eqref{eq:deformation}.
The chirally symmetric deformation 
$V_{+}$ 
has many practical and conceptual advantages over $V_{-}$, 
but it is not currently known how to introduce 
$V_+$
without reintroducing a sign problem. 
Nonetheless, we analyze both potentials
$V_\pm (\ol \psi, \psi)$
to explore the general workings of deformations in large 
$N_c$ 
orbifold equivalences.

The undeformed theory described by 
$\cL_{SO}$
in Eq.~\eqref{eq:SOLagrangian}
is free of sign problems. 
This can be easily demonstrated using the $C\gamma_{5}$ conjugacy relation, 
\begin{equation} \label{eq:Conjugate}
C \gamma_5 \cD ( C \gamma_5)^{-1} = \cD^*, 
\end{equation}
where 
$\cD = \gamma_\mu D_\mu + m + \mu \gamma_4$
is the Dirac operator. 
This or similar such conjugacy relations hold for gauge theories with fermions in real representations.
Unfortunately, 
the deformations we consider are not fermion bilinears, 
and one must integrate in auxiliary fields so that the deformed theory can be simulated using lattice techniques.

The symmetry properties of the Dirac operator in the 
\emph{deformed}
theory depend on the details of the deformation, 
and the way in which one introduces auxiliary fields.  
Utilizing a Fierz transformation on the chirally non-symmetric deformation, 
$V_-$, 
flavor-singlet auxiliary fields can be introduced, 
and lead to a deformed 
Dirac operator, 
$\cD_-$, 
obeying the relation 
\begin{equation}
C \mathcal{D}_- C^{-1} = -\mathcal{D}_-^{*}
,\end{equation} 
so long as 
$m=0$~\cite{Cherman:2010jj}.  
The quark mass term 
$m \bar{\psi}\psi$ 
breaks the 
$C$ 
conjugation relation of the deformed Dirac operator.   
In Ref.~\cite{Cherman:2010jj}, 
these observations were used to argue that one can avoid the sign problem in the $V_{-}$-deformed theory in the chiral limit.
It may be possible, 
however,  
that different implementations of the auxiliary fields will allow finite density lattice simulations of the $V_{+}$-deformed theory, or simulations with
generic values of the quark mass. 
We leave the exploration of such alternatives to future work.

As a final comment, 
one may need to add additional deformations as the chemical potential is increased
beyond the range of the low-energy EFT. 
For example, 
for large enough chemical potentials, 
$\mu \sim m_\rho$, 
it may be necessary to add a repulsive interaction between quarks coupled in the vector diquark channel. 
In this way, 
one can prevent 
$b \rho$'s 
from condensing, 
and so on. 
Whether higher spin bmesons condense is a question that cannot be addressed within the EFT, 
and for this reason we do not consider this possibility further here.

\section{Orbifold equivalence of the deformed $SO(2N_{c})$ theory and QCD} %
\label{sec:HadronsAndOrbifolds}                   						        %

The existing proofs of the necessary and sufficient conditions for orbifold equivalences involve some rather intricate analysis, 
using either lattice loop equations~%
\cite{Kovtun:2003hr} 
or large $N_{c}$ coherent state methods~%
\cite{Kovtun:2004bz},  
and there are some subtleties in applying the existing proofs directly to the orbifold projection relating the $SO(2N_{c})$ 
 theory and QCD.   Here we give an argument that the $SO(2N_{c})$ 
 theory is large-$N_{c}$ equivalent to QCD using nothing more than standard results on large $N_c$ hadron phenomenology, which we review below.  As will be clear below, the argument is not rigorous enough to be called a proof, but we believe that it is (at the least) highly suggestive and gives a nice heuristic picture of the workings of large $N_{c}$ equivalence.  
 
First, let us briefly review the key implications of the 't Hooft large $N_{c}$ limit for the behavior of hadrons in confining non-Abelian gauge theories~\cite{tHooft:1973jz,Witten:1979kh}.   In the 't Hooft large $N_{c}$ limit, the number of flavors is fixed as $N_{c}\rightarrow \infty$.  By simple large-$N_{c}$ counting, one can show that large $N_{c}$ gauge theories with quarks have an infinite number of stable meson states, which do not mix with glueballs, which are also stable.  We will mostly ignore glueballs in what follows for simplicity, which is a sensible thing to do at large $N_{c}$ thanks to the suppression of glueball-meson mixing.  By meson we mean a color-singlet state with two valence fermions;  we make no separate assumption on whether or not one of the valence fermions is an antiquark.  Moreover, one can show that the three-meson interaction vertex must scale as $N_{c}^{-1/2}$, a four-meson vertex must scale as $1/N_{c}$, and so on, while the matrix element for a current to create a meson from the vacuum scales as $N_{c}^{1/2}$.  These scalings hold regardless of whether the gauge group is unitary, orthogonal, or symplectic~\cite{Cicuta:1982fu,Lovelace:1982hz}. 
So mesons are stable and weakly-interacting at large $N_{c}$.    

The implication of these results is that confining large $N_{c}$ gauge theories are essentially classical field theories of weakly-interacting stable mesons~\cite{WittenMasterField,ColemanAspectsofSymmetry}.  We will refer to such field theories as `master field theories.'  Since there are an infinite number of mesons at large $N_{c}$, these master field theories have an infinite number of $n$-meson coupling constants $f_{n,[m]}= c_{[n,m]} N_{c}^{1-n/2}$, where $[m]$ labels which mesons are involved in the interaction and $c_{n,[m]}$ is an $N_{c}$-independent parameter which is determined by the strong dynamics of the gauge theory.  Any scattering amplitude involving mesons can be computed in terms of these coupling constants.   In practice, since there is no known way to sum the planar diagrams for generic large-$N_{c}$ theories, the coupling constants $f_{n,[m]}$ are unknown.

\subsection{Orbifold Equivalence}
\label{sec:EquivalencePicture}

Consider the action of orbifold projections from the point of view of the large $N_{c}$ master field theory associated with the $SO(2N_{c})$ gauge theory.  Suppose that a large $N_{c}$ gauge theory has some discrete global symmetry 
$\mathbb{Z}_{\Gamma}$ 
under which some of its mesons are charged. Call this the mother theory.  
The orbifold-daughter master field theory is defined by discarding all of the mesons with non-trivial charges under $\mathbb{Z}_{\Gamma}$.   
For us, the mother theory is the 
$SO(2 N_c)$ 
theory.  The relevant discrete global symmetry is the subgroup of $U(1)_{B}$ generated by $\omega = e^{i \pi/2}$, 
which acts as a 
$\mathbb{Z}_{2}$ 
symmetry on its excitations having the quantum numbers of two valence quarks.   
The neutral mesons of this theory are just the states referred to generically as mesons in Sec.~\ref{sec:OrbifoldReview}, and couple to operators of the form $\bar{\psi} \cdots \psi$, while the charged mesons couple to operators of the form $\psi^{T}C \cdots \psi$ and were called bmesons earlier.  At the level of the large $N_{c}$ master field theory, the projection to the orbifold daughter theory simply consists of discarding the bmesons.  The daughter theory contains only $U(1)_{B}$-neutral mesons.  Below we will argue that the orbifold daughter theory is large-$N_{c}$ equivalent to the mother theory.

Of course, 
we would like to identify the projected theory with QCD~\cite{Cherman:2010jj}.  
This is the difficult step in an argument restricted to the hadronic level, 
as we cannot look at the microscopic Lagrangian.\footnote{We are indebted to Masanori Hanada for very useful discussions on this point.}  
We started with an 
$SO(2N_{c})$ 
gauge theory, 
which is just a classical field theory of mesons and bmesons at large 
$N_{c}$. 
The orbifold daughter master field theory contains only 
$\mathbb{Z}_{2}$ 
neutral mesons.  
\emph{If} we assume that the daughter master field theory in fact comes from a large $N_{c}$ gauge theory, it is extremely plausible that it comes from large $N_{c}$ QCD, which 
contains only $U(1)_{B}$-neutral mesons, in contrast to gauge theories with orthogonal and symplectic gauge groups.  However, we do not know how to show that the orbifold daughter field theory arises from a large $N_{c}$ gauge theory using purely hadronic-level arguments at large $N_{c}$.  
The reason this issue may arise is the following.  From the perturbative arguments for large-$N_{c}$ equivalence, 
we know the projection symmetry of the 
$SO(2N_{c})$ 
theory must be embedded in color-space as described in Sec.~\ref{sec:HadronsAndOrbifolds} for the orbifold-daughter theory to be identified as large $N_{c}$ QCD.  However, since mesons and bmesons are color-singlets, this refinement of the projection symmetry is not visible at the level of the large $N_{c}$ master field theory.\footnote{It is conceivable that one might be able to show the necessity of choosing the correct embedding of the projection symmetry into the color group by demanding that the orbifold-daughter theory remains consistent at finite $N_{c}$ by carefully examining $1/N_{c}$ corrections, but we leave an exploration of this to future work.}   Presumably the daughter master field theory arising from a projection that is not appropriately embedded in color space does \emph{not} arise from a large $N_{c}$ gauge theory, while the daughter master field theory arising from a projection appropriately embedded in $U(1)_{B}$ and $SO(2N_{c})$ symmetries does arise from a large $N_{c}$ gauge theory.  

In what follows, 
we will assume that the projection symmetry defining the daughter theory has been appropriately embedded in color space,  so that the orbifold-daughter theory is large $N_{c}$ QCD.  We now argue that the parent and daughter theories must have the same neutral-meson correlation functions so long as the $\mathbb{Z}_{2}$ symmetry is not spontaneously broken.  

\begin{figure}[tbp]
	\centering
	\includegraphics[width=.9\columnwidth]{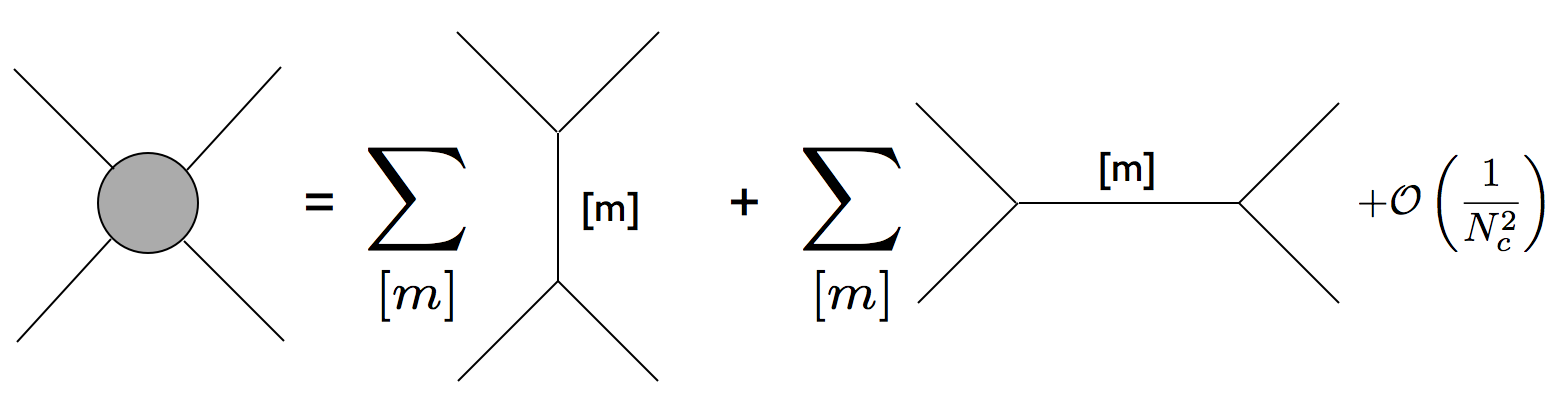}
	\caption{
	In the 't Hooft large $N_{c}$ limit, two-to-two meson scattering amplitudes are given by the sum of tree-level diagrams.  If the external legs of these tree diagrams are neutral under a global $\mathbb{Z}_{2}$ symmetry, and this symmetry is not spontaneously broken, only $\mathbb{Z}_{2}$-neutral mesons can appear as internal lines in the tree diagrams.  This observation underlies the argument for large $N_{c}$ equivalence discussed in the text.
	}
	\label{fig:LargeNTree}
\end{figure}

Consider the two-to-two scattering amplitude of neutral mesons, which is non-trivial in both the mother and daughter theories.  (The extension to generic scattering amplitudes will be obvious.)  In the mother theory, the leading contribution to the scattering amplitude scales as $1/N_{c}$ and is given by the sum of all tree-level diagrams with neutral-meson external legs, as shown in Fig~\ref{fig:LargeNTree}.   In the daughter theory, the leading contribution to the scattering amplitude is also from tree-level diagrams with neutral-meson external legs.  

Since there are no $\mathbb{Z}_{2}$-charged mesons in the daughter theory, 
only neutral mesons can appear can appear inside the tree diagrams 
contributing to the scattering amplitude.  
On the other hand, 
the mother theory does contain charged mesons, 
and if they were to appear inside the tree-level mesonic diagrams contributing to the scattering amplitude, 
the amplitudes in the mother and daughter theories would not be the same.  
Fortunately, 
so long as the external legs in the mother theory are not charged under 
$\mathbb{Z}_{2}$, 
the internal lines in tree diagrams in the mother theory must \emph{also} be neutral.  
Charged internal lines would violate 
$\mathbb{Z}_{2}$-charge conservation, 
and would be inconsistent with our assumption that 
$\mathbb{Z}_{2}$ 
is not spontaneously broken.  
Thus all of the tree-level mesonic diagrams contributing to the 
two-to-two 
scattering amplitude in the parent and daughter theory will coincide.   In both theories, the scattering amplitude will depend on the neutral-meson coupling constants entering the relevant tree diagrams.  Since the daughter theory inherits its meson coupling constants from the mother theory by construction, the scattering amplitude as a whole must coincide in the two theories.  The same conclusions will obviously hold for generic scattering amplitudes involving neutral external legs.  Thus we see that the mother and daughter theories are large-$N_{c}$ equivalent.

Of course, it is not hard to extend the hadronic picture of orbifold equivalence to the scattering amplitudes of glueballs.  The reason that we have been able to ignore glueballs in the discussion above is that glueball-meson mixing is suppressed in the large-$N_{c}$ limit.  For instance, the glueball-meson-meson coupling constant scales as $1/N_{c}$, while the glueball-glueball-meson-meson vertex scale as $1/N_{c}^{2}$~\cite{Witten:1979kh}.  So tree-level hadronic processes involving glueball internal lines are suppressed relative to the observables we discussed above.   Meanwhile, all diagrams in the $SO(2N_{c})$ theory with glueball external legs have non-trivial images in the daughter theory. This is simply because glueballs are created by color-singlet glue operators, none of which is annihilated by the projection.%
\footnote{A subtle point is that while all glueball correlation functions in the $SO(2N_{c})$ theory have corresponding correlators in QCD, there are glueball correlation functions in QCD which do not match to anything in the $SO(2N_{c})$ theory, because there are charge-conjugation odd glueballs in $SU(N_c)$ gauge theories but not in $SO(2 N_c)$ gauge theories.  We thank L. Yaffe and M. \"Unsal for discussions on these issues. }

\begin{figure}[tbp]
	\centering
	\includegraphics[width=.9\columnwidth]{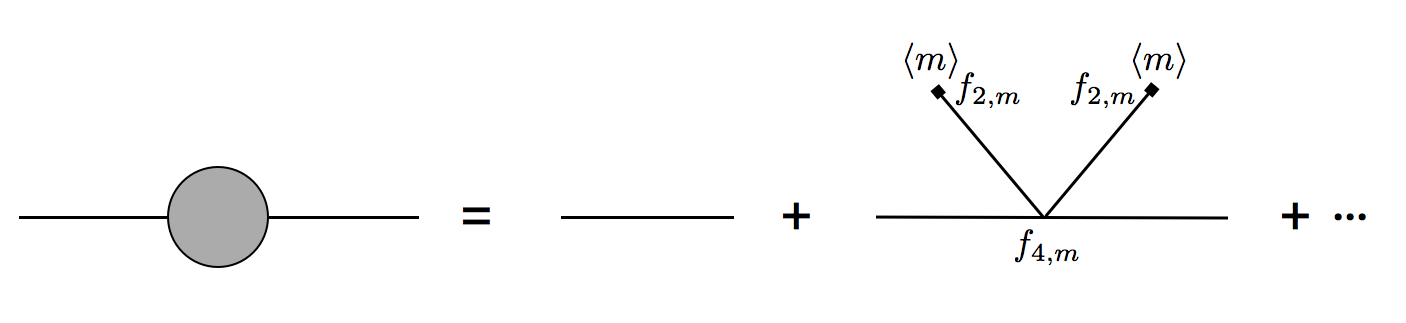}
	\caption{	Some contributions to a neutral meson propagator in the presence of a charged meson condensate. 
	The first diagram on the right is just the tree-level meson propagator.  The second diagram on the right represents a contribution from the condensate:  there is a 4-meson vertex $f_{4,m} \sim N_{c}^{-1}$ and two couplings to the charged condensate $\langle m \rangle \sim N_{c}^{1}$, each scaling as $f_{2,m} \langle m \rangle \sim N_{c}^{1/2} $.  As a result the rightmost diagram, which includes couplings to the condensate, scales the same way as the the bare neutral meson propagator.  This leads to a shift in the masses of the neutral mesons in the mother theory relative to the meson masses in the daughter theory.}
	\label{fig:MesonPropagatorWithCondensate}
\end{figure}

Now let us see how the equivalence can fail.  The crucial assumption in the argument above was that $\mathbb{Z}_{2}$ charge is a conserved charge.  If the $\mathbb{Z}_{2}$ symmetry is spontaneously broken, that assumption will be violated, and the two theories will disagree at leading order in the $1/N_{c}$ expansion.  Most obviously, our arguments that $\mathbb{Z}_{2}$-charged internal legs cannot appear in diagrams contributing to tree-level scattering amplitudes cannot hold if the $\mathbb{Z}_{2}$ symmetry is broken.  As another illustration of the violence wreaked on large $N_{c}$ equivalence by $\mathbb{Z}_{2}$ symmetry breaking, consider a neutral meson propagator in the mother theory.  The tree-level propagator in the mother theory will receive corrections from charged meson loops, while there are no such contributions in the daughter theory.  This difference, however, is suppressed by powers of $1/N_c$.   Suppose now that the $\mathbb{Z}_{2}$ symmetry breaks.  At large $N_{c}$,  the order parameters for the symmetry breaking will be the expectation values of the charged meson fields.
Because the charged mesons are color singlets, 
the order parameters for $\mathbb{Z}_{2}$ symmetry breaking will scale as $N_{c}^{1}$.  Figure~\ref{fig:MesonPropagatorWithCondensate} gives an example of the effects of such condensates:  the meson propagator in the mother theory now receives \emph{leading-order} contributions from interactions with the condensate.  This shifts the mass of the neutral mesons in the mother theory relative to the masses of the mesons in the daughter theory. So $\mathbb{Z}_{2}$-symmetry breaking destroys the equivalence.

The hadronic picture of orbifold equivalence presented above illustrates the necessary and sufficient conditions for equivalence found in \cite{Kovtun:2003hr,Kovtun:2004bz}:  in order for two theories to be orbifold equivalent in the large $N_{c}$ limit, they must be (i) related by an orbifold projection, and (ii)  the symmetry used in the orbifold projection must not be spontaneously broken.

\subsection{Deformations}
It is possible to deform a mother theory in ways that affect the properties of charged mesons but not those of the neutral mesons, at least in the absence of symmetry breaking.  Perhaps the simplest example of a deformation is the addition of a chemical potential 
$\mu$ 
for the 
$\mathbb{Z}_{2}$ 
charge.  
On physical grounds, we know that such a chemical potential shifts the masses of the charged mesons, 
but does not otherwise affect the theory unless 
$\mu$ 
is large enough to make the lightest of the charged mesons condense, 
breaking 
$\mathbb{Z}_{2}$.%
\footnote{Strictly speaking, the statement that nothing happens to neutral mesons at a small enough $\mu$ is only true at zero temperature, $T$.  In the large $N_{c}$ limit, however, large-$N_{c}$ volume independence implies that finite-$T$ effects are suppressed until the theory goes through a phase transition~\cite{Cohen:2004cd}.}    
So long as the $\mathbb{Z}_{2}$ symmetry is unbroken, shifts in the masses of the charged mesons do not affect any of the arguments for equivalence we gave above:  the scattering amplitudes of neutral mesons will continue to match in the mother and daughter theories.

Let us now turn to more general deformations.  Consider the response of the mother theory to the addition (by hand) of new charged meson vertices.  We refer to the addition of these new vertices as `the deformation'.  The daughter theory does not contain the new vertices by construction, since the orbifold projection simply discards all of the charged mesons.  The deformation obviously affects the scattering amplitudes of charged mesons at leading order.   However, the \emph{neutral} meson scattering amplitudes cannot be affected by the deformation to leading order in the $1/N_{c}$ expansion.  This is because charged mesons only contribute to the neutral-meson scattering amplitudes through loops, which are suppressed at large $N_{c}$.

The utility of this observation, which is in a sense merely a rephrasing and generalization of the arguments in Ref.~\cite{Unsal:2008ch}, is that one can quite generically deform large $N_{c}$ gauge theories without affecting their orbifold daughter theories.  For instance, one can try to delay the onset of orbifold-symmetry-breaking phase transitions, which destroy equivalences, by using deformations to shift the masses of the condensing modes, without affecting the neutral-sector physics.  Hence deformations can be used to expand the range of validity of orbifold equivalences.

Of course, the discussion above is at the heuristic level.  What one really wants is an understanding of how a specific deformation of the microscopic Lagrangian affects a large $N_{c}$ master field theory.  In general, this is a tall order.  However, at low energies, where we can use effective field theory methods to derive precisely the form of the large $N_{c}$ master field theory,
we will see exactly how this works for the two deformations
$V_{\pm} (\ol \psi, \psi )$
of 
$SO(2N_c)$ 
gauge theory.  The EFT analysis allows us to systematically derive the new charged-meson vertices that are induced by $V_{\pm} (\ol \psi, \psi )$.  As one might expect from the general arguments in this section, both deformations have no effect on the neutral sector to leading order in $1/N_{c}$ so long as the 
$\mathbb{Z}_{2}$ symmetry defining the neutral sector is unbroken.

\section{Low energy dynamics of the $SO(2N_{c})$ theory}
\label{sec:LowEnergyAnalysis}
In this section, we work out the low-energy effective theory (which is simply chiral perturbation theory) for the $SO(2N_{c})$ gauge theory.  After discussing the global symmetries of the deformed $SO(2N_{c})$ theory in Sec.~\ref{sec:Symmetries} and the pattern of spontaneous symmetry breaking in Sec.~\ref{sec:SpontaneousSymmetryBreaking}, we construct the low-energy theory in Sec.~\ref{sec:ChiPT}.  

\subsection{Symmetries of the $SO(2N_{c})$ theory}
\label{sec:Symmetries}

In the absence of the deformation, 
the theory described by 
Eq.~\eqref{eq:L}
possesses a well-known enhanced flavor symmetry, 
which can be made manifest by introducing so-called conjugate quarks, 
$\tilde{\psi}_R$. 
Starting with the usual decomposition of the Dirac spinor $\psi$ into left- and right-handed Weyl spinors, 
$\psi = (\psi_{L}, \psi_{R})^{T}$, 
one can define $\tilde{\psi}_{R}$ as $\psi_R = - \sigma_2 \tilde{\psi}_R^*$, so that%
\footnote{
There is actually a $U(1)$ ambiguity in the definition of the conjugate quark, since we could have defined it as
$\psi_R = e^{ i \phi} \sigma_2 \tilde{\psi}_R^*$, 
with 
$\phi$
arbitrary. 
The physics is independent of the angle
$\phi$. 
}
\begin{equation} \label{eq:conjugate}
\psi= 
\begin{pmatrix}
\psi_L \\
- \sigma_2 \tilde{\psi}_R^*
\end{pmatrix}
.\end{equation}
After some algebra, 
we can rewrite the action density in a simplified form with the help of the definition 
\begin{equation} \label{eq:bigpsi}
\Psi
= 
\begin{pmatrix}
\psi_L \\
\tilde{\psi}_R
\end{pmatrix}
.\end{equation}
When written in terms of $\Psi$ the Lagrangian Eq.~\eqref{eq:L} takes the form
\begin{equation} \label{eq:reL}
\cL
=
\sum_{a =1}^{N_f}
\Psi^\dagger_a 
\left(
i \sigma_\mu D_\mu
+
\mu T_3
\right)
\Psi_a
-
\sum_{a =1}^{N_f}
\frac{1}{2} m
\left[
\Psi_a^T  \sigma_2 \, T_1 \Psi_a
+
\Psi_a^\dagger  \sigma_2  \, T_1 (\Psi^\dagger_a)^T 
\right]
+ 
V( \Psi^\dagger, \Psi),
\end{equation}
where we employed the notation 
$T_i$ 
for the Pauli matrices acting on chiral indices and 
$\sigma^{\mu} = (\vec{\sigma}, - i)$ 
is the usual Weyl vector in Euclidean space.  
In this form, 
it is obvious that the classical action has an  
$SU(2 N_f) \times U(1)$ 
symmetry when  
$m = \mu = \mathfrak{C} = 0$.  
The 
$U(1)$ 
phase symmetry corresponds to the 
$U(1)_A$ 
symmetry, 
which is broken by the chiral anomaly.  
However, 
the chiral anomaly is suppressed in the 't Hooft large $N_{c}$ limit, so will treat the 
$U(1)_{A}$ symmetry on the same footing as the other symmetries of the theory.   With 
$\mu \neq 0$ 
and 
$m= \mathfrak{C} = 0$, 
the baryon chemical potential reduces the global symmetry to a chiral symmetry, 
$U(N_f)_L \times U(N_f)_R$.   
On the other hand,  with  $m \neq 0$ and $\mu = \mathfrak{C} = 0$,  the action possesses an  $SO(2 N_f)$  symmetry.
\footnote{ 
To expose the 
$SO(2 N_f)$ 
symmetry of the mass term, 
one can use a transformation 
$F 
=
\exp \left(- \frac{i \pi}{3 \sqrt{3}} \left[ T_1 + T_2 + T_3 \right] \right)
\in U(2 N_f)$.
The matrix 
$F$ 
has the property that 
$F^T  \,i T_1 F = 1$. 
Thus by redefining the fermion fields,
$\Psi = F \Psi^\prime$, 
the mass term becomes
$\frac{i}{2}
m
\left[
\Psi^{\prime \, T}  \sigma_2 \Psi^\prime
-
\Psi^{\prime \, \dagger}  
\sigma_2  
(\Psi^{\prime \, \dagger})^T
\right]$, 
while the kinetic term is invariant.
}

In terms of the conjugate quark, 
the chirally symmetric deformation appears as
\begin{eqnarray}
V_+
(\Psi^\dagger, \Psi )
&=&
- 2 \mathfrak{C}^{2}
\sum_{a,b}
\left[
\left(
\Psi^T_a
i \sigma_2
T_L 
\Psi_b
\right)
\left(
\Psi^\dagger_b
i \sigma_2 
T_L
\Psi^*_a
\right)
+
\left(
\Psi^T_a
i \sigma_2
T_R 
\Psi_b
\right)
\left(
\Psi^\dagger_b
i \sigma_2 
T_R
\Psi^*_a
\right)
\right],
\notag \\
\end{eqnarray}
where we defined left- and right-handed projection matrices 
$T_{L,R} = \frac{1}{2} ( 1 \pm T_3)$.
Due to the explicit appearance of these matrices, 
the deformation breaks the enhanced 
$U(2 N_f)$
flavor symmetry, 
while maintaining the 
$U(N_f)_L \times U(N_f)_R$ 
chiral symmetry.  
Lastly, 
written in terms of the conjugate quark
the chirally non-symmetric deformation,
$V_- ( \Psi^\dagger, \Psi)$,
takes the form
\begin{equation} \label{eq:deformedminus}
V_- ( \Psi^\dagger, \Psi)
=
2 \mathfrak{C}^2
\sum_{a,b}
\left[
\left( \Psi^T_a i \sigma_2 T_L \Psi_b \right)
\left( \Psi^T_a i \sigma_2 T_R \Psi_b \right)
+
\left( \Psi^\dagger_b i \sigma_2 T_L \Psi^*_a \right)
\left( \Psi^\dagger_b i \sigma_2 T_R \Psi^*_a \right)
\right]
.\end{equation}
Specifying the unbroken subgroup of the flavor symmetry in the chiral non-symmetrically deformed theory is somewhat subtle. 
The deformation breaks chiral symmetry, 
and the unbroken subgroup of the flavor symmetry is na\"ively only 
$U(N_{f})_{V} \times \mathbb{Z}_{4}$, where the discrete factor is the unbroken subgroup of $U(1)_{A}$.  
However,  
it turns out that the breaking of 
$U(N_{f})_{L} \times U(N_{f})_{R}$ 
by the deformation is suppressed in the large $N_{c}$ limit.  
We will argue this below in Sec.~\ref{sec:TreeLevelSpectrum}. 
As far as the behavior of the large $N_{c}$ theory is concerned, 
both deformations actually preserve a 
$U(N_{f})_{L}\times U(N_{f})_{R}$ 
subgroup of the 
$U(2N_{f})$ flavor symmetry.

\subsection{Spontaneous Chiral Symmetry Breaking}  %
\label{sec:SpontaneousSymmetryBreaking}

Having detailed the symmetries of the deformed action, 
we take up the pattern of spontaneous symmetry breaking. 
According to Coleman and Witten
~\cite{Coleman:1980mx}, 
spontaneous symmetry breaking will occur in the undeformed theory with 
$\ol \psi \psi$ 
picking a vacuum expectation value. 
The chiral condensate has the same symmetries as the mass term of the action;
hence, 
at 
$\mu = \mathfrak{C} = 0$, 
we have the symmetry breaking pattern,
$U( 2 N_f) \to SO( 2 N_f)$.
The construction of the relevant low-energy effective field theory was first discussed in~\cite{Peskin:1980gc}.%
\footnote{
The pattern of spontaneous symmetry breaking is identical to that of 
$SU( N_c)$ gauge theory with matter in the adjoint representation, 
see~\cite{Kogut:2000ek}. 
}
The coset field, 
$\Sigma \in U ( 2 N_f ) / SO (2 N_f)$, 
is schematically written as
\begin{equation}
\Sigma_{AB}
\sim
\Psi_{A\a} \Psi_{B \b}^T  ( -i \sigma_2)_{\b \a},
\end{equation}
where the indices $A,B=1,\ldots, 2N_{f}$ run over both flavor and $L,R$ indices, with the vacuum configuration of 
$\Sigma$ 
denoted by 
$\Sigma_0$. 
Adding a small quark mass, 
$m$, 
and subsequently taking the vanishing mass limit aligns the vacuum in the direction
$\Sigma_0 = - i T_1$.

Under an 
$SU( 2 N_f)$ 
transformation 
$U$ 
of the fermion field, 
$\Psi \rightarrow U \Psi$, 
the coset field 
$\Sigma$ 
transforms as 
$\Sigma \to U \Sigma \, U^T$.  Because the vacuum is invariant under
$SO(2 N_f)$
rotations, 
we must have
\begin{equation} \label{eq:relationT}
t^i \Sigma_0 + \Sigma_0 (t^i)^T = 0
\end{equation}
for $N_{f}(2N_{f}-1)$ of the generators of $SU(2N_{f})$. This condition specifies which generators of $SU( 2 N_f)$ generate the particular $SO ( 2 N_f)$ subgroup of $SU(2N_{f})$ that leaves the vacuum invariant. 
We denote the remaining $N_f (2 N_f + 1) - 1$ generators by $X^i$.  These generators satisfy the relation
\footnote{
The relation in Eq.~\eqref{eq:relationX} can be easily derived by working in the primed basis, 
namely with 
$\Psi^\prime = F^\dagger \Psi$, where the vacuum configuration of the coset field 
$\Sigma^\prime$
is 
$\Sigma_0^\prime = 1$, 
the 
$t^{\prime \, i}$
are generators of 
$SO ( 2 N_f)$, 
and 
$X^{ \prime \, i}$ 
are the symmetric traceless generators of 
$SU( 2 N_f)$. 
Reverting to the unprimed basis, 
we have
$F X^{\prime \, i} F^\dagger = X^i$, 
whence
$\Sigma_0 (X^i)^T = F X^{\prime \, i} F^\dagger = X^i \Sigma_0$.
}
\begin{equation} \label{eq:relationX}
X^i \, \Sigma_0 - \Sigma_0 (X^i)^T = 0
.\end{equation}
Fluctuations away from the vacuum configuration are generated by
\begin{equation} \label{eq:U}
U = \exp \left( \frac{i \eta' }{2  F_{\Pi} \sqrt{N_f}} \right)  \exp \left( \frac{ i \Pi }{ 2 F_\Pi} \right)
,\end{equation}
with 
$\eta'$
and 
$\Pi = \Pi^i X^i$
parameterizing the Nambu-Goldstone modes.
The most general traceless Hermitian matrix satisfying 
Eq.~\eqref{eq:relationX} has the form
\begin{equation}
\Pi 
=
\begin{pmatrix}
\pi &  \delta \\
\delta^\dagger  & \pi^T
\end{pmatrix}
,\end{equation}
where 
$\pi$ 
itself is traceless and Hermitian, 
i.e.~$\pi \in SU(N_f)$, 
and 
$\delta$ 
is a symmetric matrix.
These fluctuations take the vacuum configuration from
$\Sigma_0$
to 
\begin{equation}
\Sigma = U \Sigma_0 U^T
.\end{equation}
In light of Eq.~\eqref{eq:relationX}, 
we have
$\Sigma = V \Sigma_0$, 
with 
$V \equiv U^2$.

The transformation properties of the quark fields 
$\psi$
under the discrete symmetries,
$C$, 
$P$, 
$T$, 
and flavor singlet symmetries are listed in Table~\ref{t:symm}. 
From these transformations, 
we can deduce the transformations of 
$\Psi$ 
using Eq.~\eqref{eq:bigpsi}.
Consequently one can derive the transformations of $\Sigma$ and of $\Pi$ and $\eta^\prime$.  
These are also listed in the table,  
and enable us to identify interpolating operators with the same quantum numbers as the Nambu-Goldstone modes.   
We find
\begin{eqnarray}
\pi, \, \eta^\prime 
\sim
\ol \psi \gamma_5 \psi, 
\quad
\text{and}
\quad
\d 
\sim
\psi^T C \gamma_5 \psi .
\end{eqnarray}
The $\pi$ modes are pions in the usual sense,  which is not surprising since the familiar chiral symmetry breaking pattern from QCD is contained as a subgroup of the symmetry breaking pattern considered here:  
$ SU(N_f)_L \times SU(N_f)_R \subset SU( 2 N_f) \longrightarrow SU(N_f)_V \subset SO(2 N_f)$.  
The 
$\eta^\prime$
is the flavor-singlet pseudoscalar mode, 
as the notation suggests. 
Finally, 
the 
$\delta$ 
modes are not something one sees in QCD:  
they couple to scalar diquark operators.  
In contrast to the situation in 
$SU(N_{c})$ 
gauge theories, 
here the gauge group is 
$SO(2N_{c})$ 
and diquark operators are color \emph{singlets}.   
So the 
$\d$ 
are scalar Nambu-Goldstone modes carrying baryon number. 
These modes are the baryonic pions that we refer to as bpions for short.

\begin{table}
\caption{%
Symmetry transformations of the quarks and Goldstone modes. 
For $T$ transformations, 
we must treat
$\psi^T$
as
$(\psi^\dagger)^*$. 
}
\label{t:symm}
\smallskip
\smallskip
\smallskip
\centering
\begin{tabular}{|c|cc|cc|ccc|}
& 
$\psi$
&
$\quad \Psi \quad$ 
& 
$\quad \Sigma \quad $ 
& 
$\quad \quad \Pi \quad \quad $ 
& 
$\quad \pi \quad$ 
&
$\quad \eta^\prime \quad$
& 
$\quad \delta \quad $ 
\\
\hline
$C$  
&
$\gamma_2 \gamma_4 \ol \psi \, {}^T$
&
$T_2 \Psi$
&
$- T_2 \Sigma T_2$
&
$T_2 \Pi T_2$
&
$\phantom{-}\pi^T$
&
$\phantom{-}\eta^\prime$
&
$- \delta^*$
\\
$P$
&
$\gamma_4 \psi$
&
$- i \sigma_2 T_2 \Psi^*$
&
$T_2 \Sigma^\dagger T_2$
&
$- T_3 \Pi T_3$
&
$- \pi$
&
$- \eta^\prime$
&
$\d$
\\
$T$
&
$\gamma_4 \gamma_5 \psi$
&
$\sigma_2 T_1 \Psi^*$
&
$- T_1 \Sigma^\dagger T_1$
&
$-\Pi$
&
$- \pi$
&
$- \eta^\prime$
&
$- \d$
\\
\hline
$U(1)_B$
&
$e^{ i \theta} \psi$
&
$e^{ i \theta T_3} \Psi$
&
$e^{ i \theta T_3} \Sigma e^{ i \theta T_3}$
&
$e^{ i \theta T_3} \Pi e^{ - i \theta T_3}$
&
$\phantom{-}\pi$
&
$\phantom{-}\eta^\prime$
&
$e^{ 2 i \theta} \d$
\\

$U(1)_{A}$
&
$e^{ i \gamma_{5} \a} \psi$
&
$e^{ i \a} \Psi$
&
$e^{ 2 i \a} \Sigma $
&
$\Pi$
&
$\phantom{-}\pi$
&
$e^{ 2 i \a}\eta^\prime$
&
$\phantom{-} \d$
\\
      \end{tabular}
\end{table}
%
%

\subsection{Low Energy Effective Theory}     %
\label{sec:ChiPT}

Now that we have catalogued the pattern of spontaneous and explicit symmetry breaking, 
we are in a position to construct chiral perturbation theory for the $SO(2N_{c})$ gauge theory. 
Under a 
$U( 2 N_f)$
transformation
$\mathcal{U}$, 
the coset field 
$\Sigma$
has the transformation
$\Sigma 
\longrightarrow 
\mathcal{U} 
\, \Sigma \, 
\mathcal{U}^T$.
To account for the mass term in the action, 
Eq.~\eqref{eq:reL}, 
we define
$\cM = m \Sigma_0^\dagger$, 
and promote 
$\cM$ 
to a field with a spurious transformation under
$U ( 2 N_f)$, 
namely
$\cM  
\longrightarrow
\mathcal{U}^*
\cM
\,
\mathcal{U}^\dagger$.
This spurious transformation is chosen to render the mass term invariant under $U(2N_{f})$. 
In the effective theory, 
we form $U(2N_{f})$-invariant combinations with the building blocks 
$\Sigma$ 
and
$\cM$. 
When the spurion field 
$\cM$
takes on its constant value, 
symmetries will be broken in precisely the correct way.

To include the chemical potential, 
we gauge an external real-valued vector field, 
$B_\mu$~\cite{Kogut:1999iv}. 
This ensures that Ward identities are properly respected, 
and exactly fixes coefficients in the low-energy theory. 
At the day's end,
the field takes on the constant 
(imaginary) 
value
$B_\mu = - i \mu T_3 \delta_{\mu 4}$.
Under a local 
$U( 2 N_f)$
transformation, 
\begin{equation}
B_\mu 
\longrightarrow
\mathcal{U} B_\mu \mathcal{U} ^\dagger
- 
i \, \mathcal{U} \, \partial_\mu \mathcal{U}^\dagger
.\end{equation}
Derivatives involving the 
$\Sigma$
field are made covariant with the definition
\begin{eqnarray}
D_\mu \Sigma
&=&
\partial_\mu \Sigma
+ 
i B_\mu \Sigma
+ 
i \Sigma B_\mu^T
.\end{eqnarray}

The effective Lagrangian is constructed from spurion fields by demanding invariance under 
$U( 2 N_f)$
transformations. 
Additionally we require invariance under the $C$, $P$, and $T$ transformations in Table~\ref{t:symm}.  
To leading order in the symmetry breaking parameters
$\mu$ 
and 
$m$, 
we have
\footnote{
With 
$\Sigma$ 
living in 
$U(2N_{f})/SO(2N_{f})$ 
rather than 
$SU(2N_{f})/SO(2N_{f})$, 
we can form three additional invariants: 
$\frac{F_{\Pi}^{2}}{N_{c}} |\ln \det \Sigma|^{2}$,
$\det \left( \Sigma \cM
+ 
\Sigma^\dagger \cM^\dagger \right)$, and
$\det \left( D_\mu \Sigma D_\mu \Sigma^\dagger \right)$. 
The first captures the contribution of the chiral anomaly to the $\eta'$ mass~\cite{Witten:1980sp}, and is $1/N_{c}$ suppressed relative to the terms shown in Eq.~\eqref{eq:EFTnoDeformation}.  The second invariant is of order
$m^{N_f}$, 
and so only appears at leading order in a one-flavor theory. 
The last invariant gives rise to a difference between the decay constant of the 
$\eta^\prime$, 
denoted
$F_{\eta^\prime}$, 
and the remainder of the Nambu-Goldstone modes, which is an effect that is suppressed at large $N_{c}$~\cite{Kaiser:2000gs}.  
Hence we do not consider these additional invariant operators. 
}
\begin{equation}
\label{eq:EFTnoDeformation}
\cL
=
\frac{F_\Pi^2}{4}
\tr
\left[
D_\mu \Sigma D_\mu \Sigma^\dagger
\right]
-
\frac{\lambda F_\Pi^2}{4}
\tr
\left[
\Sigma \cM
+ 
\Sigma^\dagger \cM^\dagger
\right]
.\end{equation}
Matching to the properties of mesons expected in large $N_{c}$ gauge theories, the low-energy constants 
$F_{\Pi}$ 
and 
$\lambda$
appearing above must scale as  
$F_{\Pi}^{2} \sim N_{c}^{1}$, 
and 
$\lambda\sim N_{c}^{0}$.  
This Lagrangian captures the low-energy physics of the undeformed theory.   So Eq.~\eqref{eq:EFTnoDeformation} is the low-energy limit of the large $N_{c}$ master field theory discussed in Sec.~\ref{sec:HadronsAndOrbifolds}.

We must also add the effects from the deformations, 
$V_\pm (\Psi^\dagger, \Psi)$. 
Four-quark operators have long been treated in chiral perturbation theory~\cite{Bernard:1985wf}. 
The spurion method provides an efficient way to account for four-quark operators.\footnote{ 
See, for example, applications of chiral perturbation theory to lattice discretization effects~\cite{Bar:2003mh,Tiburzi:2005vy}. }
We assume the power counting 
$\mathfrak{C}^2 \sim \mu^2 \sim m$
so that leading order constitutes one insertion of the deformation. 
To match the deformation onto the EFT using spurions, 
it is useful to rewrite the deformation potential, 
Eq.~\eqref{eq:pmdeformations}, 
in an ornate form.
Working first with the chirally symmetric deformation $V_{+}$, 
we write
\begin{eqnarray}
V_+( \Psi^\dagger, \Psi)
= 
-2 \mathfrak{C}^2
\sum_{a,b=1}^{N_{f}}
\left[
\left(
\Psi^T
i \sigma_2
T_L^{(ab)} 
\Psi
\right)
\left(
\Psi^\dagger
i \sigma_2 
T_L^{(ba)}
\Psi^*
\right)
+
\left(
\Psi^T
i \sigma_2
T_R^{(ab)} 
\Psi
\right)
\left(
\Psi^\dagger
i \sigma_2 
T_R^{(ba)}
\Psi^*
\right)
\right],
\notag \\
\end{eqnarray} 
where the flavor structure is now contained entirely in the matrices
$T^{(ab)}_{L,R} = T_{L,R} \, \lambda^{(ab)}$, 
with 
$(\lambda^{ab})_{cd} = \d^a_c \d^b_d$. 
The fixed matrices 
$T^{(ab)}_{L,R}$
are now promoted to spurions
$L^{(ab)}$ and $R^{(ab)}$, 
which transform in the same manner
\begin{eqnarray}
L^{(ab)} 
&\longrightarrow&
\mathcal{U}^* L^{(ab)} \, \mathcal{U}^\dagger
\notag \\
R^{(ab)}
&\longrightarrow&
\mathcal{U}^* R^{(ab)} \, \mathcal{U}^\dagger
.\end{eqnarray} 
The matrices 
$T^{(ba)}_{L,R}$, 
on the other hand,
are promoted to the Hermitian conjugate spurions
$L^{(ab)  \dagger}$
and
$R^{(ab) \dagger}$, 
respectively.

Now we can map 
$V_+( \Psi^\dagger, \Psi)$ 
into the low-energy theory. 
At leading order, 
we only have 
$\Sigma$ 
and
$\Sigma^\dagger$
fields at our disposal, 
i.e.~no derivatives, and no quark-mass insertions. 
For the 
$L^{(ab)} \otimes L^{(ab) \dagger}$
operator appearing in 
$V_+(\Psi^\dagger, \Psi)$, 
there is just one invariant that can be formed
\footnote{
When the spurions take on their constant values, 
operators coupling specially to the singlet field
$\eta^\prime$, 
such as 
$\det [ \Sigma L^{(ab)} ] \tr [ \Sigma^\dagger L^{(ab) \dagger}]$,
vanish. 
}
$\tr [ \Sigma L^{(ab)} ] \tr [ \Sigma^\dagger L^{(ab) \dagger}]$.
For the 
$R^{(ab)} \otimes R^{(ab) \dagger}$
operator, 
there is an analogous invariant. 
The coefficient of both terms must be the same because $V_{+}$
is invariant under the interchange
$\{ L \leftrightarrow R \}$. 
To maintain invariance under flavor, 
moreover, 
each flavor combination must be identically weighted, 
and thereby we find
\begin{equation}
V_+^{\text{EFT}} 
=
c_{+} \, F_\Pi^2
\sum_{a,b = 1}^{N_{f}}
\Big(
\tr \left[ \Sigma L^{(ab)} \right]
\tr \left[ \Sigma^\dagger L^{(ab) \dagger} \right]
+
\tr \left[ \Sigma R^{(ab)} \right]
\tr \left[ \Sigma^\dagger R^{(ab) \dagger} \right]
\Big)
.\end{equation}
Here 
$c_{+}$ 
is a new low-energy constant
with mass dimension two, 
and must be directly proportional to
$\mathfrak{C}^2$, 
with the factor of 
$F_\Pi^2$
chosen for convenience.

For the chirally non-symmetric deformation $V_{-}$, 
from Eq.~\eqref{eq:deformedminus}
we have
\begin{equation}
V_- ( \Psi^\dagger, \Psi)
=
2 \mathfrak{C}^2
\sum_{a,b=1}^{N_{f}}
\left[
\left( \Psi^T i \sigma_2 L^{(ab)} \Psi \right)
\left( \Psi^T i \sigma_2 R^{(ab)} \Psi \right)
+
\left( \Psi^\dagger i \sigma_2 L^{(ab) \dagger} \Psi^* \right)
\left( \Psi^\dagger i \sigma_2 R^{(ab) \dagger} \Psi^* \right)
\right].
\end{equation}
For the 
$L^{(ab)} \otimes R^{(ab)}$
operator appearing above, 
there are two invariants: 
$\tr [ \Sigma L^{(ab)} ] \tr [ \Sigma R^{(ab)}]$,
and
$\tr [ \Sigma L^{(ab)} \Sigma R^{(ab)} ]$. 
Combining these invariants with their Hermitian conjugates,  
we find that there are only two terms needed in the chiral Lagrangian to account for the
$V_-$ 
deformation at this order.%
\footnote{
We note that in general the $V_{-}$ deformation induces multiplicative quark mass renormalization, while additive renormalization is forbidden because $V_{-}$ maintains a discrete chiral symmetry $\mathbb{Z}_{4} \subset U(1)_{A}$.   The effects of multiplicative renormalization show up in the EFT formalism as terms involving insertions of both the mass spurion and the deformation spurion fields.  However, such terms are  beyond the order to which we are working here.
} 
The low-energy physics of the deformed theory is described by the terms
\begin{eqnarray}
V_-^{\text{EFT}}
&=&
c_{-} F_{\Pi}^{2}
\sum_{a,b=1}^{N_{f}}
\left( 
\tr [ \Sigma L^{(ab)} ] \tr [ \Sigma R^{(ab)}] + \tr [ \Sigma^\dagger L^{(ab) \dagger} ] \tr [ \Sigma^\dagger R^{(ab) \dagger} ]
\right)
\notag \\
&& + 
d_{-} F_{\Pi}^{2}
\sum_{a,b=1}^{N_{f}}
\left(
\tr [\Sigma L^{(ab)} \Sigma R^{(ab)}] + \tr [ \Sigma^\dagger L^{(ab) \dagger} \Sigma^\dagger R^{ (ab) \dagger} ]
\right)
\,.\end{eqnarray}
The low-energy constants 
$c_{-},d_{-}$
also have mass dimension two, and must be directly proportional to 
$\mathfrak{C}^2$.

To summarize: 
at leading order in each of the symmetry breaking parameters, 
$\mathfrak{C}^2$, $\mu^2$, and $m$, 
the low-energy EFT is described by the Lagrangian
\begin{equation}
\cL_\pm 
= 
\frac{F_\Pi^2}{4}
\tr
\left[
D_\mu \Sigma D_\mu \Sigma^\dagger
\right]
-
\frac{\lambda F_\Pi^2}{4}
\tr
\left[
\Sigma \cM
+ 
\Sigma^\dagger \cM^\dagger
\right]
+ 
V_\pm^{\text{EFT}}
.\end{equation}
As a check, 
one can easily show that the terms appearing above in 
$V_\pm^{\text{EFT}}$ 
are each invariant under
$C$, 
$P$, 
and 
$T$, 
as well as 
$U( N_f)_V$
flavor transformations.   As advertised in Sec.~\ref{sec:HadronsAndOrbifolds}, the effective field theory analysis has allowed us to identify the precise form of the interactions important to the low-energy physics of the large $N_{c}$ master field theory due to deformations of the microscopic theory.

\subsection{Tree-Level Spectrum}
\label{sec:TreeLevelSpectrum}
To get some intuition about the physics that is encoded in the low-energy effective theory, 
it is helpful to examine its tree-level spectrum.  
To compute the tree-level spectrum, we need to expand 
$\Sigma$ 
to second order in the Nambu-Goldstone modes around 
$\Sigma_{0}$, 
the vacuum value of the chiral field.  
For now, 
let us suppose that we are in the vacuum specified by 
$\Sigma_{0} = -i T_{1}$, 
which corresponds to the vacuum state when
$\mu = \mathfrak{C} = 0$.
Hence we are assuming that there is no bpion condensation when the parameters 
$\mu$ 
and 
$\mathfrak{C}$ 
are non-zero.  
The computation will tell us the parameter values at which this assumption will fail, modulo an interesting subtlety discussed in Sec.~\ref{sec:ChiralOdd}.  
After getting some physical intuition from the analysis in this section, 
we will address the problem of vacuum alignment in general in Sec.~\ref{sec:VacuumOrientation}.

Upon plugging in 
$\Sigma = V \Sigma_{0} = -i V T_{1}$ 
together with the final values of the spurion fields 
into the effective Lagrangian, 
we obtain 
\begin{align}
\label{eq:EFTLagrangianV}
\mathcal{L} &= \frac{F^{2}_{\Pi}}{4} \tr\left[D_{\mu}V D_{\mu} V^{\dag}\right] 
- 
\frac{\lambda F_{\Pi}^{2}}{4} \tr \left[V m + V^{\dag}m\right] 
+
V_\pm^{\text{EFT}}(V^\dagger, V)
,\end{align}
where $V = U^{2}$, 
with $U$ given in Eq.~\eqref{eq:U}, 
and  
\begin{align}
D_{\mu} V &= \partial_{\mu}V + \mu \delta_{\mu 4} [T_{3},V] \\
D_{\mu} V^\dagger &= \partial_{\mu}V^\dagger + \mu \delta_{\mu 4} [T_{3},V^{\dag}] 
,\end{align}
The effect of the deformation appears in the potentials,
\begin{eqnarray}
V_+^{\text{EFT}} (V^\dagger, V)
&=&
c_{+} F_\Pi^2
\sum_{a,b=1}^{N_{f}}
\left(
\tr \left[ V T_{-} \lambda^{(ab)} \right]
\tr \left[ V^\dagger  T_{+} \lambda^{(ba)} \right]
+ 
\tr \left[ V T_{+} \lambda^{(ab)} \right]
\tr \left[ V^\dagger  T_{-} \lambda^{(ba)} \right]
\right)
,\notag \\ 
\label{eq:V+EFT}
\end{eqnarray}
for the chirally symmetric deformation, 
and
\begin{eqnarray}
V_-^{\text{EFT}}(V^\dagger, V)
&=& 
-
  c_{-}  F_{\Pi}^{2} \sum_{a,b=1}^{N_{f}}{ \left(
	\tr \left[ V T_{+} \lambda^{(ab)} \right]\tr \left[ V T_{-} \lambda^{(ab)} \right]   
 +
	\tr \left[ V^{\dag} T_{+} \lambda^{(ba)} \right]\tr \left[ V^{\dag}T_{-} \lambda^{(ba)} \right] 
     \right)} \notag \\
&&-
  d_{-}  F_{\Pi}^{2} \sum_{a,b=1}^{N_{f}}{ \left(
	\tr \left[ V T_{+} \lambda^{(ab)}  V T_{-} \lambda^{(ab)} \right]   
 +
	\tr \left[  V^{\dag} T_{+} \lambda^{(ba)} V^{\dag} T_{-} \lambda^{(ba)} \right] 
     \right)}     
\label{eq:V-EFT}
,\end{eqnarray}
for the chirally non-symmetric deformation. 
Appearing above are the raising and lowering matrices, 
which are given by
$T_{\pm} = \frac{1}{2}(T_{1} \pm i T_{2})$.

Expanding Eq.~\eqref{eq:EFTLagrangianV} to second order in $\Pi$ and $\eta^\prime$, 
one can read off the mass terms for the $\pi$, $\delta$, $\delta^{\dag}$, and $\eta'$ modes. 
Here mass refers to the pole masses in the pion or bpion propagators. 
Note that the commutators in $D_{\mu}$ ensure that the chemical potential does not contribute to the masses of the $\pi$ and $\eta'$ modes.  
Identifying 
$m_{\pi}^{2} = \lambda m$ 
to be the mass of the Nambu-Goldstone bosons when 
$\mu = \mathfrak{C} = 0$, 
our results for the masses are given in Table~\ref{table:Masses}.  
To interpret these results we need to know the signs of 
$c_{\pm}$ 
and 
$d_{-}$,
as well as the large 
$N_{c}$ 
scaling of these low-energy constants.

\begin{table}
\begin{center}
\smallskip
\smallskip
\smallskip
\begin{tabular}{|c|c|c|}
\hline
Mode & Mass with $V_{-}$ deformation &  Mass with $V_{+}$ deformation\\
\hline
$\pi$ & $(m_{\pi}^{2}+4 d_{-} )^{1/2}$ &  $m_{\pi}$ \\
\hline
$\eta'$ & $(m_{\pi}^{2}+4 d_{-} )^{1/2}$ &  $m_{\pi}$ \\
\hline
$\delta$  & $ (m_{\pi}^{2}+ 4 c_{-} )^{1/2}  + 2\mu$   &  $(m_{\pi}^{2}+ 4 c_{+} )^{1/2}  + 2\mu$  \\
\hline
$\delta^{\dagger}$ & $(m_{\pi}^{2}+ 4 c_{-} )^{1/2}-2\mu$ & $ (m_{\pi}^{2}+ 4 c_{+} )^{1/2} - 2\mu$  \\
\hline
  \end{tabular}
\caption{%
Tree-level Nambu-Goldstone masses in the $SO(2N_{c})$ gauge theory in a phase with unbroken 
$U(1)_{B}$ symmetry with the two deformations 
$V_{\pm}$.  
Mass refers to the pole mass in the propagator. 
At large 
$N_{c}$, 
the effect of turning on the deformations is to change the 
$\delta$ 
masses but not the 
$\pi$ 
and 
$\eta'$ 
masses, 
because 
$c_{\pm} \sim N_{c}^{0}$, 
and 
$d_{-} \sim N_{c}^{-1}$.
}  
\label{table:Masses}
\end{center}
\end{table}

\begin{figure}[tbp]
	\centering
	\includegraphics[width=.6\columnwidth]{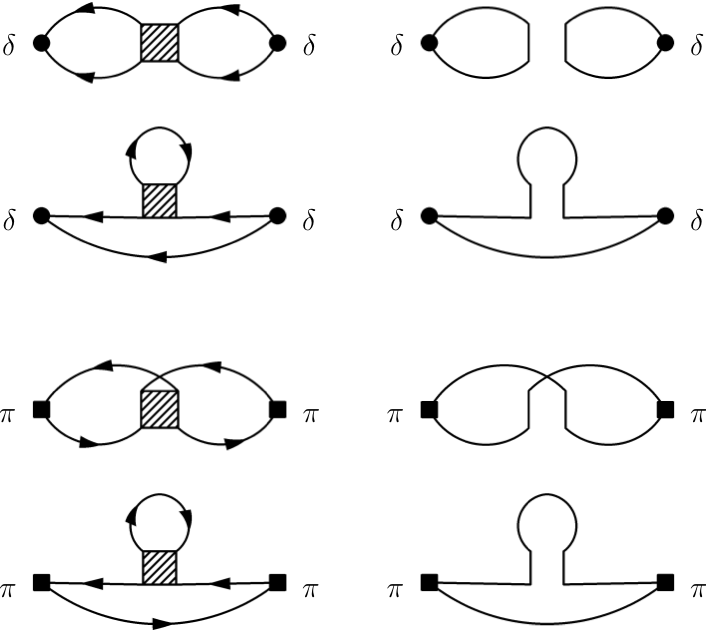}
	\caption{Matrix elements of the deformation operators 
	$V_{\pm}$ 
	between bpionic and pionic states.  
	In the diagrams on the left-hand side, which illustrate the quark contractions that contribute to the matrix elements, 
	the deformation operator is represented by a hatched box, while the lines are quarks.  The arrows track the direction of $U(1)_{B}$ charge flow.  
	Note that the deformation is always $U(1)_{B}$-charge neutral, as it must be.  On the right, we give the color-flow diagrams associated with each contraction.      
	Also note that the pionic matrix elements must vanish for the $V_{+}$ deformation operator in the chiral limit.}
	\label{fig:DeformationScaling}
\end{figure}

To fix the large 
$N_{c}$ 
scaling and the signs of the low-energy constants induced by turning on the deformation, 
one must consider the scaling of matrix elements of 
$V_{\pm}$ 
between operators with the quantum numbers of pions and bpions. 
We show the relevant diagrams in Fig.~\ref{fig:DeformationScaling}.  
From symmetry arguments, 
it is obvious that matrix elements of 
$V_{+}$ between pion states vanish in the chiral limit, 
while those of 
$V_{-}$ 
do not.  
In any case, 
the leading color contraction contributing to bpionic matrix elements has an extra power of 
$N_{c}$ 
compared to the leading color contraction involving pions.  
In the low energy theory, 
this means that we must have 
$c_{\pm} \sim N_{c}^{0}$, 
while 
$d_{-}$ is suppressed, 
scaling as 
$d_{-} \sim N_{c}^{-1}$.

Since $d_{-}$ is negligibly small at large $N_{c}$, we only need to determine the sign of 
$c_{\pm}$ 
to understand the effect of the deformations on the Nambu-Goldstone boson masses.  
To determine the sign of 
$c_{\pm}$, 
we use a QCD-inequality-like argument.  From the expressions in Table~\ref{table:Masses},  
it is clear that the effect of the deformation in the EFT is to shift the bpion masses.   
In the microscopic theory, the shift in the bpion mass due to the deformation 
$V_{\pm}$ 
is encoded in the matrix element depicted at the top of Fig.~\ref{fig:DeformationScaling}:
\begin{equation}
\mathfrak{M_{\pm}}
=
\langle S_{ab}(z) V_{\pm}(y) S^\dagger_{ab}(x) \rangle.
\end{equation}
$\mathfrak{M}_{\pm}$ is evaluated in the full \emph{undeformed} theory, which allows us to use the  the conjugacy relation in 
Eq.~\eqref{eq:Conjugate}
for 
$\cD$, 
which
is the Dirac operator of the undeformed theory.   
In the limit of large time separations only the ground-state bpions contribute to 
$\mathfrak{M}_{\pm}$.  
So the sign of 
$\mathfrak{M_{\pm}}$ 
controls whether the deformation term in the action raises or lowers the energy of a bpion state, 
and thus determines the sign of 
$c_{\pm}$.  
What we seek to show is that the sign of 
$\mathfrak{M_{\pm}}$ is controlled by the sign of 
$\mathfrak{C}^2$.  
Choosing 
$\mathfrak{C}^2>0$
then forces $c_{\pm} > 0$.

In general there are two quark contractions contributing to bpion matrix elements of the deformation, 
$\mathfrak{M_{\pm}}$. 
Only one contraction, 
however, 
contributes in the large 
$N_c$ 
limit. 
Working with a fixed background gauge field 
$A_\mu$, 
and using the 
$C\gamma_{5}$ 
conjugacy relation,
the contractions evaluate to
\begin{eqnarray}
\mathfrak{M}_{\pm}^{A}
&=&
\mathfrak{C}^2
\Bigg(
\tr \left[  G(z,y) G^\dagger (z,y) \right]
\tr \left[  G(y,x) G^\dagger (y,x) \right]
\notag \\
&& \phantom{sp} \pm
\tr \left[  G(z,y) \gamma_5 G^\dagger (z,y) \right]
\tr \left[  \gamma_5 G(y,x) G^\dagger (y,x) \right]
\Bigg)
+
\mathcal{O}(1 / N_c) ,
\end{eqnarray} 
where $G = \cD^{-1}$ is the fermion propagator in the fixed background field $A_\mu$.
The terms in the first line above are manifestly positive, 
while those in the second are real. 
Owing to the Cauchy-Schwartz inequality, 
we know that
\begin{equation}
\Big|
\tr \left[  G(z,y) \gamma_5 G^\dagger (z,y) \right]
\Big|
\leq 
\tr \left[  G(z,y) G^\dagger (z,y) \right]
,\end{equation}
and consequently 
$\mathfrak{M}_{\pm}^{A} \geq 0$.   
Since the integration measure of the undeformed theory is positive, this relation will survive integration over $A_\mu$, 
yielding 
$\mathfrak{M}_\pm \geq 0$.
Taking long time separation, 
\emph{etc}., 
we have 
$c_{\pm} \geq 0$. 
Because 
$c_{\pm}$ 
must be directly proportional to some positive power of 
$\mathfrak{C}^2$, 
taking 
$\mathfrak{C}^2>0$ 
indeed forces $c_{\pm}>0$.

The physics resulting from the deformations is now clear.  
The effect of \emph{both} deformations is to raise the bpion masses by an 
$\mathcal{O}(N_{c}^{0})$ 
amount while leaving the masses of the neutral-sector 
$\pi$ and $\eta'$ 
modes unchanged to leading order in the 
$1/N_{c}$ 
expansion.  
Note that when 
$c_{\pm} = d_{-} = 0$, 
the masses of 
$\delta^{\dagger}$ modes become tachyonic when 
$\mu > m_{\pi}/2$.  
Once this happens, 
the vacuum alignment shifts, 
and the masses of all of the modes change.  
In the presence of the 
$V_{\pm}$ 
deformations the critical value of 
$\mu$ 
becomes 
\begin{equation}
\mu^{\text{crit}}_{B}  
= \sqrt{ \left( \frac{m_{\pi}}{2} \right)^{2} + c_{\pm}}
.\end{equation} 
Because both 
$c_{\pm}$ 
are positive if $\mathfrak{C}^2>0$, 
turning on either deformation pushes the bpion condensation point past 
$m_{\pi}/2$.  
These low-energy constants, 
moreover, 
can be made larger by taking the value of 
$\mathfrak{C}^2$
to be larger.

Once $\mu$ is large enough to make the bpion masses negative, 
there will be a second-order phase transition to a bpion-condensed phase.  
The phase diagrams of the deformed theories are explored in detail in 
Sec.~\ref{sec:VacuumOrientation} below.  
In addition to the bpion-condensed phase, 
it turns out that for some values of the parameters the theory with the 
$V_{-}$ 
deformation also has an exotic phase with both bpion and 
$\eta^\prime$ 
condensation.  
Unlike the pure bpion-condensed phase, 
the exotic phase cannot be predicted from staring at the tree-level spectrum in 
Table~\ref{table:Masses} 
since it is a 
\emph{metastable} 
phase, 
separated from the other two phases by first-order phase transitions.

\section{Vacuum Orientation and the fate of $U(1)_{B}$}%
\label{sec:VacuumOrientation}

In constructing the low-energy effective theory,  we have incorporated fluctuations about the  vacuum alignment,  $\Sigma_0 = -i T_{1}$.  This alignment characterizes the ground state of the theory with all sources of explicit symmetry breaking turned off,  namely $m = \mu = \mathfrak{C} = 0$.   In Sec.~\ref{sec:TreeLevelSpectrum}, we computed the tree-level spectrum of the effective theory as a function of 
$m$, 
$\mu$, 
$c_{\pm}$, 
and 
$d_{-}$ 
under the assumption that the vacuum alignment remains 
$\Sigma_0 = -i T_{1}$ 
even when these parameters are non-zero.  
However, this is not always justified, because the theory can undergo phase transitions where the vacuum alignment changes.  
This is already evident in the tree-level spectrum described in Sec.~\ref{sec:TreeLevelSpectrum}, 
since some of the Nambu-Goldstone modes become tachyonic for certain values of the parameters.  
In the absence of a deformation, 
we will see that when 
$\mu>m_{\pi}/2$, 
the bpions condense.  
As already suggested by the tree-level spectrum analysis, 
the deformations can prevent this phase transition.  
The more complete analysis in this section reveals that the 
$V_{-}$ 
deformed theory has an additional phase not present in the undeformed theory.  
The new phase has both bpion and 
$\eta^\prime$ 
condensation, 
and is metastable.
As this exotic phase is absent for the 
$V_{+}$ 
deformed theory, 
we handle the analysis of its vacuum orientation first. 
We begin with general considerations that apply to both cases.

\subsection{General Considerations}    
\label{sec:VacuumAlignmentGeneralConsiderations}            %

In analyzing the vacuum alignment, we choose to keep  $\Sigma$
in the form
$\Sigma = V \Sigma_0$, 
with 
$\Sigma_0 = - i T_1$ 
fixed. 
Thus we are looking for the value  
$V(x) = V_0$ 
that minimizes the vacuum energy density, 
with 
$V_0 =  \mathbbm{1}$
corresponding to the vacuum orientation when
$m = \mu = \mathfrak{C} = 0$.
Without specifying the dynamics, 
the vacuum alignment is already highly constrained.
The matrix 
$V_0$
must be unitary, 
and satisfy the transposition constraint
\begin{equation} \label{eq:transposey}
V_0^T = T_1 V_0 T_1
,\end{equation}  
which follows from Eq.~\eqref{eq:relationX}.
We will 
\emph{assume} 
that the vacuum alignment does not violate 
$CPT$ for simplicity of presentation, but will take note as necessary on what happens when this assumption is relaxed.
Violation of 
$CPT$
implies Lorentz symmetry violation but not vice versa. 
The action in Eq.~\eqref{eq:L} 
does not contain explicit sources of
$CPT$
violation. 
From Table~\ref{t:symm}, 
the combination of 
$CPT$ 
transformations leads to the restriction
\begin{equation}
[ V_0, T_1 ] = 0
.\end{equation}
Combined with the transposition constraint in Eq.~\eqref{eq:transposey}, 
$CPT$
invariance implies
\begin{equation} \label{eq:symmetric}
V_0^T = V_0
.\end{equation}

Away from the large 
$N_{c}$ 
limit, 
$V_{0}$ 
would also need to satisfy 
$\det(V_{0})=1$. 
This is no longer the case at large $N_{c}$; 
the determinant of 
$V$ carries the information about the flavor-singlet 
$\eta'$ 
Nambu-Goldstone mode.  
For what follows, 
it is convenient to write 
$V_0$
as an 
$SU( 2 N_f)$ 
matrix in four 
$N_f \times N_f$-blocks  
times an overall 
$U(1)$ 
phase, 
\begin{align}
V_0 = e^{ i \varphi} \begin{pmatrix} A & B \\ C & D \end{pmatrix}.
\end{align}
Imposing the transposition constraint implies that 
$B = B^T$, 
$C = C^T$
and
$D = A^T$, 
while 
$CPT$ 
invariance requires
$B = C$. 
Unitarity of $V_{0}$ then gives us the relations
\begin{eqnarray} \label{eq:unitary}
A A^\dagger 
+ 
B B^* 
&=& 
\mathbbm{1}, 
\notag \\
A B^* 
+ 
B A^*
&=& 
0
.\end{eqnarray}

With the symmetric condition in Eq.~\eqref{eq:symmetric} imposed, 
$V_0$
is constrained only up to conjugation by some matrix 
$W \in SO( 2 N_f)$,
in the form
$V_0 \to W V_0 W^T$. 
Subsequent imposition of the transposition constraint in
Eq.~\eqref{eq:transposey}
forces the matrix 
$W$
to satisfy
$[ T_1, W ]  = 0$.
This commutation constraint implies that 
$W$ 
lives in an
$SO(N_f) \times SO(N_f)$
subgroup of 
$SO ( 2 N_f)$.
The remaining constraints on 
$V_0$
arise from the dynamics.

\subsection{Chirally Symmetric Deformation $V_{+}$}

Now let us consider minimization of the vacuum action in the case of the chirally symmetric deformation, 
$V_+$. 
Inserting everything known from Sec.~\ref{sec:VacuumAlignmentGeneralConsiderations} about the vacuum orientation
into Eqs.~\eqref{eq:EFTLagrangianV} and \eqref{eq:V+EFT}, 
we find the vacuum action density, 
$S_+$, 
takes an especially simple form
\begin{equation} \label{eq:Splus}
S_+
=
\frac{m_{\pi}^{2} F_\Pi^2}{2 a_+}  
\left\{
\tr 
\left[
( \cA^\dagger - a_+)
( \cA - a_+)
\right]
-
\tr 
\left[
1
+
a_+^{2}
\right]
\right\}
,\end{equation}
with the matrix 
$\cA$
given by
$\cA = A \, e^{i \varphi}$, 
the constant
$a_+$
defined as
\begin{equation} \label{eq:APlus}
a_+
=
\frac{(m_{\pi}/2)^{2}}{ \mu^2 -c_{+} }
,\end{equation}
and 
$m_{\pi}^{2} = \lambda m$ 
as above. 
For 
$a_+ > 0$, 
the global minimization of 
$S_+$
follows that considered in~\cite{Kogut:2000ek}.
Ignoring momentarily the constraints on the matrix
$\cA$, 
the trace in 
Eq.~\eqref{eq:Splus} 
gives the distance in the 
$2 N_f^2$-dimensional 
space between the complex matrix elements of 
$\cA$ 
and the real diagonal matrix 
$a_+ \mathbbm{1}$. 
When 
$0 < a_+ < 1$, 
the global minimum can be achieved with 
$\cA = a_+ \mathbbm{1}$. 
This fixes the matrix 
$B$ 
up to sign,
namely
$B e^{i \varphi} = \pm i \sqrt{ 1 - a_+^2} \, \mathbbm{1}$. 
So for $0<a_{+}<1$, 
the value of the action at the global minimum is 
\begin{align}
s_{+} \equiv \frac{S_{+}}{m_{\pi}^{2}F_{\Pi}^{2}N_{f}} = -\frac{1+a_{+}^{2}}{2a_{+}} < -1
\end{align}
On the other hand, 
when 
$a_+ > 1$, 
the distance is minimum when the hypersphere has the largest possible radius consistent with unitarity. 
This demands
$\cA = \mathbbm{1}$, 
and consequently
$B = 0$, with the value of the action at the minimum given by 
$s_{+}=-1$.

The case 
$a_+ < 0$
is a possibility unique to the deformed theory. 
For this case, 
the vacuum action still measures the distance from the constant matrix 
$a_+ \mathbbm{1}$. 
The sign of the overall pre-factor, 
however, 
requires us to 
\emph{maximize}
the distance from 
$a_+ \mathbbm{1}$. 
This is achieved by making the matrix elements of 
$\cA$
real, 
diagonal, 
and each as large as possible, 
namely
$\cA = \mathbbm{1}$.  Unitarity again forces $B=0$, and $s_{+} = -1$ at the minimum.
Thus we have found the vacuum alignment
\begin{equation}
\label{eq:AlignVplus}
V_0
= 
\begin{cases}
\mathbbm{1}, 
&
\quad
\text{for} 
\quad
a_+ < 0, 
\text{ or }
a_+ > 1
\\
a_+ \mathbbm{1} 
\pm i \sqrt{1 - a_+^2} \, T_1,
&
\quad
\text{for} 
\quad
0 < a_+ < 1
\end{cases}
.\end{equation}
Using the transformation rules in Table~\ref{t:symm}, 
we see the phase encountered when 
$0 < a_+ < 1$ 
breaks 
$C$, 
$T$, 
and
$U(1)_B$, 
while maintaining 
$P$, 
and 
$CT$.\footnote{
In addition to breaking
$U(1)_B$, 
the bpion condensed phase also breaks the 
$SO( 2 N_f)_V$
symmetry. 
The vacuum alignment in Eq.~\eqref{eq:AlignVplus} still preserves a subgroup of this vector symmetry, 
under which
$V_0 \to \mathcal{O} V_0 \mathcal{O}^T$, 
where
$\mathcal{O} = \diag \left( O, O \right)$, 
and
$O \in SO(N_f)$. 
This symmetry corresponds to the subgroup
$SO( N_f)_V$. 
Constraints on the realization of this vectorial symmetry along the lines of the Vafa-Witten theorem~\cite{Vafa:1983tf} 
are considered in Appendix~\ref{AppendixA}.
} 
This phase is precisely the bpion-condensed phase which must be avoided for the orbifold equivalence to be valid.  If we had not assumed CPT invariance, there would have been an extra phase phase in $V_{0}$ coming from the matrix $C$, and we would have seen a $U(1)$ degeneracy in the orientation of $V_{0}$.  This degeneracy is associated with the Nambu-Goldstone mode associated with the breaking of $U(1)_{B}$.  The two vacua seen in Eq.~\eqref{eq:AlignVplus} are the subset of these vacua which are invariant under CPT, in which $\langle \delta \rangle = \langle \delta^{\dag} \rangle$.   

It is also worth noting that in both phases of the theory we have $\det ( V_0 ) = 1$, implying that $U(1)_{A}$ is broken. This means that there is generically a non-zero chiral condensate in the bpion-condensed phase.  As in Refs.~\cite{Kogut:1999iv,Kogut:2000ek,Son:2000xc}, as one moves from the normal phase to the bpion-condensed phase $\Sigma_{0}$ switches from pointing in the $-i T_{1}$ direction to rotating by an angle $\cos^{-1}{(a_{+})}$ toward the $\mathbbm{1}$ direction.

Notice that without the deformation, 
$c_+ = 0$, 
the theory necessarily has a phase transition to this bpion-condensed phase at 
$\mu = m_{\pi}/2$.  
The deformation, 
however, 
allows us to avoid the phase transition.  
Because the low-energy constant
$c_{+}$ 
is positive, 
we can crank up 
$\mathfrak{C}^2$ 
in 
Eq.~(\ref{eq:pmdeformations}) 
until 
$c_{+}>\mu^{2}$. 
Beyond this point,
$a_+ < 0$
and we stay in the uncondensed phase even if 
$\mu > m_{\pi}/2$.
This is exactly as we argued in Sec.~\ref{sec:TreeLevelSpectrum}.

\subsection{Chirally Non-Symmetric Deformation $V_{-}$}
\label{sec:ChiralOdd}

The analysis of the 
$V_{-}$ 
deformation parallels that of the 
$V_{+}$ 
deformation but involves some subtleties we did not encounter above.  
With the vacuum alignment in the form
\begin{align} \label{eq:blerg}
V_0 = e^{i \varphi} \begin{pmatrix} A & B \\ B & A^T \end{pmatrix}
,\end{align}
we satisfy the transposition constraint, 
and maintain 
$CPT$
invariance by assumption. 
The vacuum energy density, 
$S_-$
then takes the form
\begin{align}
\label{eq:VminusEnergyDensity}
S_{-} 
&= 
\frac{1}{2} F_{\Pi}^{2} 
\Bigg( 
(2\mu)^{2} 
\tr
\left[
A^{\dag}A -1
\right]
 - 
 m_{\pi}^{2} \, 
 \tr
 \left[
 A^{\dag} e^{ - i \varphi} 
 +
 A \, e^{ i \varphi} 
 \right] 
  - 
 2
 c_{-} 
 \tr
 \left[ 
 (B^{*})^2  e^{ - 2 i \varphi}
 + 
 B^2 e^{ 2 i \varphi} 
 \right]
 \Bigg)
\end{align}
where we dropped 
$d_{-}$, 
because it is suppressed at large 
$N_{c}$. 
With mass-degenerate quarks, 
the form of Eq.~\eqref{eq:VminusEnergyDensity}
allows us minimize the action by first diagonalizing
the matrices 
$A$ 
and 
$B$.  
Notice the unitarity constraint, 
Eq.~\eqref{eq:unitary}, 
implies that 
$A$
is automatically diagonal if 
$B$
is, 
and vice versa. 
Hence the action and unitarity allow us to work with 
$V_0$ 
in the form of Eq.~\eqref{eq:blerg} using 
$A = \diag (a_i) $ 
and
$B = \diag (b_i)$, 
with 
$i = 1, \dots, N_f$. 
With diagonal matrices inserted into
Eq.~\eqref{eq:VminusEnergyDensity}, 
there is no coupling between flavors.
Consequently the values of 
$a_i$
and
$b_i$
that minimize the action must be the same for each flavor. 
Thus the vacuum orientation must be of the form
\begin{equation} \label{eq:NeatForm}
V_0
=
e^{ i \varphi}
\begin{pmatrix}
a e^{i \alpha} \mathbbm{1} & b e^{ i \beta} \mathbbm{1} \\
b e^{i \beta} \mathbbm{1} & a e^{i \alpha} \mathbbm{1}
\end{pmatrix}
,\end{equation}
for positive real parameters
$a$, 
and
$b$. 
Unitarity restricts the size of these parameters to be less than one. 
As a result, 
the vacuum configuration is invariant under a 
$SO( N_f )_V$
symmetry, 
under which 
$V_0 \to \cO V_0 \cO^T$, 
with 
$\cO = \diag ( O, O)$, 
and 
$O \in SO( N_f)$. 
Note that we did not call on any Vafa-Witten-like theorem in this analysis, 
and instead used the effective theory to demonstrate the preservation of this vectorial symmetry. 
The applicability of the Vafa-Witten theorem to the deformed 
$SO( 2 N_c)$ 
theory is subtle.
This is discussed further in Appendix~\ref{AppendixA}, 
where we conjecture that 
$SO(N_f)_V$
cannot break spontaneously.

Looking at Eq.~\eqref{eq:NeatForm}, 
we see that the phase
$\b$
can be measured relative to 
$\a$, 
and 
$\a$
subsequently absorbed into the overall phase
$\varphi$.
Enforcing unitarity leads us to
\begin{equation}
V_0
= 
e^{i \varphi}
\left(
a \, \mathbbm{1}
\pm
i \sqrt{1 - a^2} \,
T_1
\right)
.\end{equation}
The action density for the 
$V_-$-deformed 
theory then becomes
\begin{align}
\label{eq:VminusEnergyDensity2}
s_-( a, \varphi )
\equiv
\frac{S_{-}}{ m_\pi^2 F_{\Pi}^{2} N_f}
=
-
\left[ 
\frac{1}{2}
(x -  y \cos 2 \varphi )
( 1 -  a^2)
+ 
a \cos \varphi
\right]
,\end{align}
where we have employed the dimensionless variables
\begin{equation} \label{eq:xy}
x = \frac{\mu^2}{(m_\pi / 2)^2}, 
\quad
y = \frac{c_-}{(m_\pi / 2)^2}
.\end{equation}
The minimization of 
$s_- (a, \varphi)$ 
as a function of
$a$ 
and 
$\varphi$
is detailed in Appendix~\ref{AppendixB}. 
Note that the phase of the deformed 
$SO (2 N_c)$ 
theory which is orbifold-equivalent to large 
$N_{c}$ 
QCD has 
$a=1$
and 
$\varphi = 0$.  
This phase has unbroken 
$U(1)_{B}$
symmetry.  
If 
$a$ deviates from 
$a=1$, 
baryon number will be violated by the formation of a bpion condensate, 
and the equivalence will be invalidated. 
A non-vanishing angle 
$\varphi$
corresponds to a vacuum condensate with the quantum numbers of the 
$\eta^\prime$. 
The possibility of such condensation ocurring within the domain of validity of the EFT is a consequence of the 't Hooft large 
$N_c$ 
limit, 
where the $\eta'$ meson is light.

\begin{figure}[tbp]
	\centering
	\includegraphics[width=.45\columnwidth]{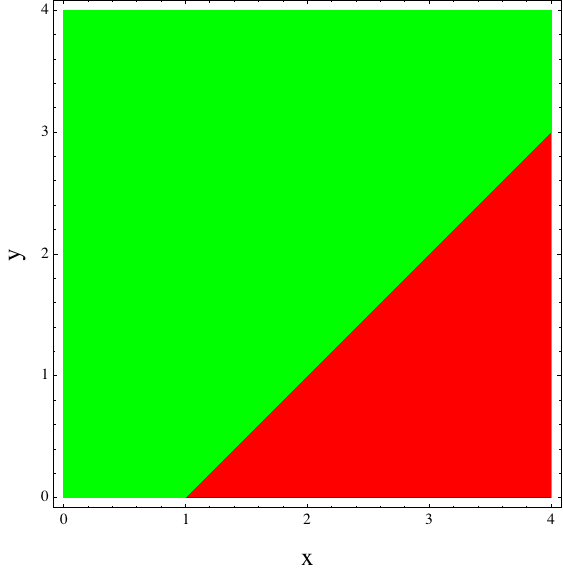}
	\includegraphics[width=.45\columnwidth]{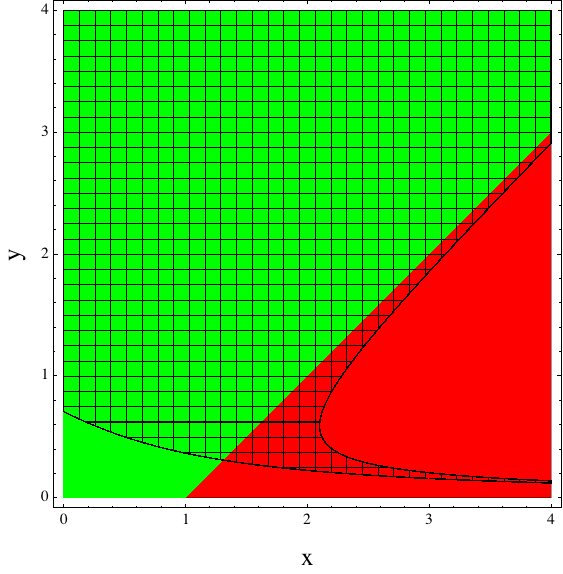}
	\caption{
	(Color online.)
	Phase diagram for the deformed 
	$SO( 2 N_c)$
	theory. 
	The figure on the left depicts the ground state of both the $V_{+}$ and $V_{-}$ deformed theories as a function of 
	$x = \mu^{2}/m_{\pi}^{2}$ 
	and 
	$y = c_{\pm}/m_{\pi}^{2}$.
	The light (green) color shows the 
	$U(1)_B$
	symmetric phase, while the dark (red) color show the bpion-condensed phase. 
	On the right, 
	we have also included the region (hatched area) in which one encounters an exotic metastable vacuum in the $V_{-}$-deformed theory with broken
	$U(1)_B$ symmetry
	and 
	$\eta^\prime$ 
	condensation. 
	In both plots, 
	the undeformed theory lives strictly on the $x$-axis. 	
	}
	\label{fig:PhaseDiagram}
\end{figure}

Defining the quantity $a_{-}$ by
\begin{equation} \label{eq:AMinus}
a_- = \frac{1}{x - y} = \frac{( m_\pi / 2)^2}{\mu^2 - c_-}
,\end{equation}
analogously to 
$a_+$ 
in Eq.~\eqref{eq:APlus},
and minimizing $s_-(a, \varphi)$ as described in Appendix~\ref{AppendixB}, we find the vacuum alignment to be given by
\begin{align} 
\label{eq:AlignVminus}
V_0
= 
\begin{cases}
\mathbbm{1}, 
&
\quad
\text{for} 
\quad
a_- < 0, 
\text{ or }
a_- > 1
\\
a_- \mathbbm{1} 
\pm i \sqrt{1 - a_-^2} \, T_1,
&
\quad
\text{for} 
\quad
0 < a_- < 1
\\
e^{i \varphi_c} 
\left(
a_c  
\mathbbm{1} 
\pm i \sqrt{1 - a_c^2} \, T_1
\right),
&
\quad
\text{for}
\quad
\text{see Figure~\ref{fig:PhaseDiagram}.}
\end{cases}
\end{align}
Comparing Eq.~\eqref{eq:AlignVminus} to Eq.~\eqref{eq:AlignVplus}, 
the phase structure is very similar with both deformations.  There a normal phase for $y>x-1$ in which the deformed 
$SO ( 2 N_c)$
theory is equivalent to QCD, and a bpion-condensed phase for $y\le x-1$ where the equivalence breaks down.

The crucial difference, 
however, 
is the existence of the phase characterized by the parameters
$a_c$
and
$\varphi_c$, 
which depend on $x$ and $y$ and are defined in Eq.~\eqref{eq:Ac}.  
In this exotic phase
there is an $\eta'$ condensate along with a bpion condensate, 
and as is apparent in Eq.~\eqref{eq:AlignVminus} there are two degenerate vacua. 
Non-vanishing values for $\eta^\prime$ and bpion condensates break not only 
$U(1)_B$ and $U(1)_{A}$,
but also
$C$, 
$P$, 
and
$T$ 
individually, 
as well as the product
$CP$.
By assumption, 
$CPT$
is maintained.  As noted in the discussion of the $V_{+}$ deformation, without this assumption the degeneracy in the location of the bpion-condensed vacua would be enhanced from $\mathbb{Z}_{2}$ to U(1).

Whether the  exotic phase exists depends rather complicatedly on the parameters
$x$ 
and 
$y$. 
When it can exist,  
the exotic phase is nevertheless metastable, 
because its vacuum energy,
$s_{-}(a_c, \varphi_c)$, 
is greater than in the other two phases, as is shown in Eq.~\eqref{eq:ExoticPhaseEnergy}. 
Because it is associated with a higher vacuum energy, 
this phase corresponds to a local minimum separated from the global minimum by a first-order phase transition.

The determination of the region in parameter space where the exotic phase exists is messy, and hence relegated to Appendix~\ref{AppendixB}.  Here we simply summarize the results of this analysis.  For the metastable phase to coexist, 
there is a minimum value for the size of the deformation,
$y_{\text{min}}$,
required, which varies with the chemical potential.   This bound comes from the need to maintain the relation $a_{c}<1$.   Of course, it must also be the case that $\cos \varphi_{c} <1$.  As a result, for a given size of the deformation there is a maximum value of the chemical potential $x_{\text{max}}$ beyond which the exotic phase does not exist.   These considerations imply that the metastable exotic phase lives between the two curves in the $(x,y)$ plane given by
\begin{align}
 x_{\text{max}} &= \frac{4 y^{2} +1 + \sqrt{16 y^{2}+1}}{4 y}\\
y_{\text{min}} &=  \frac{- x + \sqrt{x^2 + 2} }{2}.
\end{align}   
We show the phase diagram in Figure~\ref{fig:PhaseDiagram}. 
Notice that for a given value of chemical potential
$x$, 
the ground state can remain in the 
$U(1)_B$
symmetric phase for a suitable value of the deformation, 
$y$, 
namely
$y > x - 1$. 
This is merely the condition we found above:
$a_- > 1$. 
An even larger value of the deformation, 
such that
$y > x$, 
will result in 
$a_- < 0$. 
Even though the parameter
$a_-$ 
is singular at 
$x=y$, 
the theory remains in the 
$U(1)_B$
symmetric phase on either side.

To get some intuition about these results it is useful to consider  a few simple limits. 
Below we consider the metastable phase of the theory when the scale of the deformation is either small or large 
compared to the other scales in the problem.  
We also consider what happens in the chiral limit. 
In all cases, 
we assume that the parameters remain small compared to the cut-off scale of the EFT.

\begin{enumerate}
\item
Small deformation: $y  \ll x$.

In this limit, 
the region where the metastable phase exists gets pinched to a region that shrinks with 
$x$.
The minimum value for 
$y$ 
to enter the region of the metastable phase becomes
$y_{\text{min}} = \frac{1}{2 x} - \frac{1}{4 x^3} + \cO ( x^{-5} )$. 
Enforcing 
$
\cos \varphi_c < 1$
in this limit, 
we find the maximal value for 
$y$
is
$y_{\text{max}} = \frac{1}{2 x} + \frac{3}{4x^3} +  \cO( x^{-5})$.  
This pinching effect is clearly visible in Fig.~\ref{fig:PhaseDiagram}.

\item
Large deformation: $y \gg 1$.

In this region, 
$y_{\text{min}}$,
has been exceeded, 
and we must wonder whether the metastable phase persists for all values of 
$y \gg x$. 
For small 
$x$, 
we have
\begin{equation}
\cos^2 \varphi_c = \frac{1}{2} \left( 1 - \frac{1}{\sqrt{2} y} \right) + \cO(x)
,\end{equation} 
which meets the constraint
$\cos \varphi_c < 1$. 
Thus the metastable phase persists for all 
$y \gg x$, 
as Fig.~\ref{fig:PhaseDiagram} shows.

\item
Chiral limit: $x, y \gg 1$.

When the quark mass is taken to zero, 
both parameters
$x$
and
$y$
become large simultaneously, 
while their ratio, 
$\xi \equiv \frac{y}{x} = \frac{c_{-}}{\mu^{2}}$,
is a free parameter. 
In terms of 
$\xi$ and $x$, 
we have
\begin{align}
a_{c}^{2} &= 1-\frac{1}{x\sqrt{2 \xi (\xi+1)}}\\ 
\cos^2 \varphi_c 
&= 
\frac{a_c^2}{2}
\left( 1 + \frac{1}{\xi} \right) 
\end{align}
Written in this form, 
the chiral limit consists of taking $x\gg 1$.  
Clearly 
$0<a_{c}<1$ 
is always satisfied at large 
$x$,
but 
the constraint on the angle,
$\cos^{2} \varphi_c < 1$, 
is only met for
$\xi > 1$.
So in the chiral limit, the metastable phase exists so long as   
$c_{-} > \mu^{2}$.
The sign of
$\cO( x^{-1})$
corrections, 
however, 
allows 
smaller values of 
$\xi$ 
to satisfy 
$\cos \varphi_c < 1$
as the chiral limit is approached. 
This feature can also be seen in Figure~\ref{fig:PhaseDiagram}, 
as the metastable region extends slightly below the line
$y = x-1$ 
for large 
$x$.

\end{enumerate}

\section{Comparison to QCD and Conclusions}   %
\label{sec:ComparisonToQCD}                              %

Taking into account everything we have learned about the low-energy dynamics of the 
$SO(2 N_c)$
gauge theory, 
it is easy to see how the orbifold equivalence of the 
$SO(2N_c)$ 
theory with QCD 
exposes itself in the effective theory.  
At low energies, the correlation functions of 
$U(1)_{B}$ 
neutral operators will be describable within the effective theory.  
These correlation functions simply encode the scattering of 
$\pi$
and 
$\eta'$ 
modes.  
So long as 
$U(1)_{B}$ 
is unbroken, such scattering amplitudes computed in the 
$SO( 2 N_c)$ 
effective theory will be the same as ones computed with an effective theory with the 
$\delta$ 
and 
$\delta^\dagger$
modes deleted.  
We now claim that if the bpion modes are deleted from the coset field
$\Sigma$, 
then Eq.~\eqref{eq:EFTLagrangianV} simply describes the usual chiral perturbation theory for large $N_{c}$ QCD.  
This is obviously the case in the undeformed theory.  For instance, there is no coupling of $\Sigma$ to $\mu$ when the charged 
$\delta$ 
and
$\delta^{\dagger}$ 
modes are deleted, 
as must be the case because the pions are neutral under 
$U(1)_{B}$.  
In the deformed theory, 
things are somewhat more subtle, as we now explain;
but, 
the conclusion is the same.

When either of the deformations we considered is turned on, 
it is relatively simple to see how the theory remains large-$N_{c}$ equivalent to QCD at low energy.  
Essentially, the deformation terms do not appear in the projected theory to leading order in the $1/N_{c}$ expansion, as expected from the general arguments of 
Sec.~\ref{sec:OrbifoldReview} and Sec.~\ref{sec:HadronsAndOrbifolds}.  
As discussed above, 
the $V_{-}$ deformation shifts the 
$\pi$ and
$\eta'$ 
masses by a 
$1/N_{c}$ 
suppressed amount relative to the undeformed theory, 
while the $V_{+}$ deformation does not shift the $\pi$ and $\eta'$ masses at all. 
It is also straightforward to verify that both deformations do not introduce any new interactions at tree level for pions and 
$\eta'$ mesons to leading order.%
\footnote{Unlike the $V_+$ deformation,  
however,
the $V_{-}$ deformation does induce $1/N_{c}$-suppressed pion interactions.}    
Reassuringly, the $V_{\pm}$ deformations have precisely the same effect on the neutral sector of the deformed theory:  
in net, there is no effect at large $N_c$ so long as $U(1)_{B}$ is unbroken.
While the two deformations affect correlation functions involving 
bpion modes differently,
they both alter the masses of the 
$\delta$ and $\delta^{\dagger}$ modes, 
pushing the onset of bpion condensation away from 
$\mu = m_{\pi}/2$.  
This implies that one can keep the deformed 
$SO(2N_c)$ 
theories in a phase with unbroken 
$U(1)_{B}$ 
symmetry even once 
$\mu\geq m_{\pi}/2$ 
by cranking up the coefficients of the deformation without affecting neutral-sector physics.  
As a result, the deformed 
$SO(2 N_c)$ 
theories remain large $N_{c}$ equivalent to QCD past 
$\mu = m_{\pi}/2$.

We can also see the failure of the equivalence in the 
$U(1)_{B}$-broken phases.  
In the bpion-condensed phase, 
the pions and bpion modes mix with each other at leading order, 
and the physical modes appropriate to the condensed phase do not map onto anything in QCD.  
For instance, 
in the condensed phase there is a Nambu-Goldstone boson associated with the spontaneous breaking of
$U(1)_{B}$, which is massless even in the deformed theories, since there is no explicit $U(1)_{B}$ breaking in these theories. 
When the 
$SO( 2 N_c)$
theory is in a 
$U(1)_{B}$ broken phase, 
it is not large 
$N_{c}$-equivalent to QCD.

Let us now address one of the main issues motivating our analysis.  
The chirally non-symmetric deformation, 
$V_{-}$, 
is attractive from a practical point of view. 
A way to implement this deformation on the lattice has been devised that is devoid of a sign problem in the chiral limit, 
even at finite 
$\mu$~\cite{Cherman:2010jj}.   
Looking at Eq.~\eqref{eq:pmdeformations},
however, 
one may worry that even though 
$V_{-}$ 
penalizes  
$U(1)_{B}$-breaking bpion condensation, 
it simultaneously seems to subsidize 
$U(1)_{B}$-breaking condensation that violates parity.  
Of course, 
the lightest pseudoscalar bmesons, 
which are the most obvious candidates for modes whose condensation would break parity and 
$U(1)_{B}$, 
should have masses much larger than the pion mass,
while the lightest scalar bmesons are pseudo-Nambu-Goldstone bosons with vanishing mass in the chiral limit.  
As a result, 
one might have expected that small deformations would prevent scalar 
$U(1)_{B}$-breaking condensation without triggering parity-breaking condensation.

Given these considerations, 
it is perhaps somewhat surprising that the effective field theory analysis in Sec.~\ref{sec:VacuumOrientation} revealed the $V_{-}$-deformed theory does indeed have a parity and $U(1)_{B}$-breaking vacuum when the deformation is large enough compared to the chemical potential, but still small compared to the chiral 
symmetry breaking scale.
Rather than being associated with condensed pseudoscalar bmesons, which are not included in the effective theory, this exotic parity-broken phase contains condensed bpions and $\eta'$ mesons.  Fortunately, however, the parity-broken phase is always metastable where it exists at all.  The true ground state of the theory, at least so long as we are within the domain of validity of the EFT, is always parity-conserving, with the realization of the $U(1)_{B}$ symmetry determined by the relative sizes of the pion mass
$m_{\pi}^{2}$, 
chemical potential
$\mu^{2}$, 
and the deformation coefficient
$\mathfrak{C}^2$.

It is also interesting to note that there is no hint within the EFT that the parity-broken phase ever becomes competitive in energy compared with the parity-unbroken phase when the deformation is large.\footnote{
We can make this conclusion just from the tree-level analysis, without having to think about meson loop corrections, 
because meson loops are suppressed in the large $N_{c}$ limit.
}    In terms of  the normalized vacuum energy 
$s_{-} (a, \varphi)$ 
introduced in Sec.~\ref{sec:VacuumOrientation}, when $c_{-} \gg \mu^{2}$, $s_{-}(a_c, \varphi_c) \to -1/\sqrt{2}$, while $s_{-}(1,0) = -1$.   More generally, 
we were able to show analytically that
$s_-(a_c, \varphi_c) > s_-(1,0)$, 
when
$c_{-} > \mu^{2}$.
This inequality 
indicates that the $U(1)_{B}$-preserving vacuum remains the stable one.  
Of course, none of this excludes the possibility that once the deformation and the chemical potential become large enough, 
the pseudoscalar bmesons might condense.  
In this regime, 
however, 
the EFT arguments do not apply. 
Equally well, 
one cannot be sure that pseudoscalar bmeson condensation will happen just by contemplating Eq.~\eqref{eq:pmdeformations}, 
which can only be expected to control the effective potential for $\psi^{T} C \psi$ when the deformation is small.  
Once the deformation coefficient becomes large (that is, $\mathfrak{C}^2  \gtrsim \Lambda_{SO(2N_c)}^{2}$), one should expect large quantum corrections to the effective potential for $\psi^{T} C \psi$, 
and the locations of the minima of this potential depend on non-perturbative physics that can only be studied using lattice simulations.

We close by noting a few open problems and directions for future work.  
It seems important to search for a way to implement the chirally symmetric deformation 
$V_{+}$ without a sign problem at finite $\mu$.  
While we have not been able to find such an implementation thus far, 
we also have not been able to find a no-go argument.  
The existence of a sign-problem-free but somewhat baroque implementation of auxiliary fields that enables Monte Carlo simulations of 
$V_{-}$ 
in the chiral limit hints that the sign problem in the 
$SO(2 N_c)$ 
theory might be 
just
a technical problem that can be dodged if one is  sufficiently stubborn.  
Finding a sign-problem-free implementation of the 
$V_{+}$-deformed theory would be especially nice because it is much cleaner theoretically than 
$V_{-}$, 
in the sense of having a simpler phase structure and no breaking of chiral symmetry (even $1/N_{c}$ suppressed).

Another important problem is to understand the mapping between baryonic observables between the $SO(2N_{c})$ theory and large $N_{c}$ QCD.  
Baryon interpolating operators are $\mathbb{Z}_{2}$-neutral when $N_{c}$ is even, 
suggesting that baryons may be in the common sector of the two theories; 
but, 
the details of applying the orbifold equivalence to operators involving color-space epsilon tensors have not yet been worked out.  
It would also be nice to develop a sharp understanding of the conditions under which the equivalence persists in the Veneziano large 
$N_{c}$ limit, 
where meson loops become unsuppressed.

Lattice simulations of the 
$V_-$-deformed theory near the chiral limit would be rather ambitious. 
Even though dramatic progress has recently been made in simulating QCD
at (or near) the physical light quark masses, 
it is unclear how light the quark masses need to be in the current context
to enable Monte Carlo simulations of the deformed
$SO( 2 N_c)$
theory without causing a sign problem. 
To this end, 
it would be useful to find an implementation of the auxiliary fields which does not suffer from a sign problem at non-vanishing quark masses.

Finally there is the difficult issue of including irrelevant operators in the continuum limit. 
If one takes the continuum limit of the deformed theory na\"ively, 
then the deformation terms will scale as 
$\sim a^2$, 
where 
$a$ is the lattice spacing, and their effects will be suppressed as we send $a\rightarrow 0$. 
One needs to take the continuum limit in such a way that the effects of the deformation terms remain large enough to prevent bpion condensation. To do this, it will be necessary to input the lattice deformation coefficient as
$\mathfrak{C}_{\text{latt}}^{2} \sim \mathfrak{C}^2  / a^{2}$, and then fine-tune $\mathfrak{C}^{2}_{\text{latt}}$ to produce low-energy coefficients $c_{\pm}$ of the desired size. 
In doing this one must keep in mind that the na\"ive scaling with the lattice spacing
is in fact modified by radiative corrections, 
and one must additionally account for operator mixing. 
This issue is especially acute for the $V_-$-deformed theory, 
for which chiral symmetry is broken; 
and, 
one must consider the full compliment of four-quark operators. 
These problems can be tackled using lattice perturbation theory.

While there are many hurdles that currently need to be overcome in order to simulate the deformed 
$SO( 2 N_c)$ 
theory at finite density, 
we hope nevertheless that this work stimulates new activity and insight into the very difficult problem of finite density QCD.

\acknowledgments{%
We are indebted to Masanori Hanada for collaboration at the initial stages of this work, discussions on related projects, and for innumerable useful conversations.  We are also very grateful for  the hospitality of the TQHN group at the University of Maryland, College Park, where this work was initiated, and thank Paulo Bedaque and Tom Cohen for discussions and spiritual guidance at the initial stages of this work.  A.~Cherman is grateful to Adi Armoni, Mike Buchoff, Simon Hands, Mithat \"Unsal, Matt Wingate and Larry Yaffe for useful discussions.
B.~C.~Tiburzi is grateful to  J.-W.~Chen, W.~Detmold, F.-J.~Jiang, C.-J.~D.~Lin, and B.~Smigielski for useful conversations, 
and to the Taiwan National Center for Theoretical Sciences (North) for their hospitality and partial support.
This work is supported by
the U.~S.~Department of Energy, Office of Nuclear Physics, under grant number
DE-FG$02$-$94$ER$40818$ (B.C.T.). 
}

\appendix

\section{Spontaneous breaking of vector symmetries}   %
\label{AppendixA} 						              %

Here we remark on the constraints posed by the famous theorems of Vafa and Witten~\cite{Vafa:1983tf,Vafa:1984xg} 
on the impossibility of spontaneous breaking of parity and vector-like symmetries for the $SO(2N_{c})$ gauge theory.  First, we note that  the theorem regarding vector-like symmetries does not apply at finite density:  with non-vanishing chemical potential, 
vector symmetries can be spontaneously broken.  
For example, 
baryon number can be spontaneously broken by diquark condensation.%
\footnote{
What has been shown in theories where fermions are in real representations, 
however, 
is that vector symmetries cannot break spontaneously provided that 
$\mu < m_\pi / 2$, 
see~\cite{Aloisio:2000rb}.
}
Furthermore, the applicability of both theorems to theories deformed by four-quark operators is quite subtle, 
as was noted in Ref.~\cite{Vafa:1983tf} and is well-known in the context of irrelevant operators in lattice QCD with Wilson fermions, 
where the parity and isospin-breaking Aoki phase can appear at finite lattice spacing for some values of the parameters of the lattice 
action~\cite{Aoki:1983qi,Sharpe:1998xm}.   
Thus the existence of the bpion-condensed and $\eta'$ phases in the deformed 
$SO(2N_{c})$ 
theory is not in contradiction with the Vafa-Witten theorems.

While the above remarks imply that we cannot easily call on the Vafa-Witten theorems in analyzing the 
$SO(2N_{c})$ 
theory, 
in this appendix we conjecture that 
$SO(N_f)_V$
symmetries
cannot be broken in the deformed theories. 
The argument in this appendix is not called on in the body of the paper; 
because, 
as will be clear below, 
it involves some plausible but hard to prove assumptions.  
The analysis in the main text of the paper is, 
however, 
entirely consistent with the conjecture that 
$SO(N_{f})_{V}$ 
symmetries cannot be spontaneously broken in the deformed 
$SO(2N_{c})$ 
theories.
In this appendix, 
we consider flavor breaking condensates, 
much like the preliminary considerations in the classic work by Vafa and Witten%
~\cite{Vafa:1983tf}.

To consider diquark condensation, 
we must add diquark sources to the action in Eq.~\eqref{eq:L}.
In a background gauge field, 
the action is 
\begin{equation} \label{eq:sourced}
\cL 
=
\sum_{a =1}^{N_f}
\ol \psi_a 
\cD (m_a)
\psi_a 
+
\sum_{a=1}^{N_f} 
\left( 
J_a S_{aa}  
+ 
J^\dagger_a S_{aa}^\dagger
\right)
.\end{equation}
In writing 
$\cL$, 
we assume that the requisite auxiliary fields to handle the deformation have been integrated in, 
for example, as in%
~\cite{Cherman:2010jj},
The operator
$\cD(m_a)$
contains the gauged kinetic term, 
chemical potential, 
mass term, 
and all terms with auxiliary fields. 
Our considerations extend trivially to the undeformed theory merely by dropping the auxiliary field terms. 
The mass dependence has been explicitly shown, 
as we now allow for non-degenerate masses. 

It is convenient to introduce Nambu-Gor'kov fields%
~\cite{Morrison:1998ud,Hands:1999zv}. 
With 
$\Psi_{NG} 
= \left(
\begin{smallmatrix} 
\psi
\\ 
\phantom{.} \ol \psi {}^T
\end{smallmatrix}\right)$, 
and
$\ol \Psi_{NG}
= 
\left( 
\psi^T, 
\ol \psi \, 
\right)$, 
we can write the action in the form
\begin{equation}
\cL 
= 
\sum_{a=1}^{N_f}
\ol \Psi_{a, NG} \, K (J_a, m_a) \Psi_{a,NG}
,\end{equation}
where 
$K (J_a, m_a)$ appears as
\begin{equation}
K (J_a, m_a)
= 
\begin{pmatrix}
J_a C \gamma_5 
& 
\frac{1}{2} \cD_a
\\
- \frac{1}{2} \cD_a^T
&
- J_a^\dagger C \gamma_5
\end{pmatrix}
,\end{equation}
and satisfies
$K (J_a, m_a)^T = - K (J_a, m_a)$, 
with all flavor dependence explicitly labeled. 
The antisymmetry allows us to perform the Gaussian integration over the Nambu-Gor'kov fields producing 
$\Pf [ K (J_a, m_a) ]$. 

In this fixed background of gauge and auxiliary fields, 
the chiral and diquark condensates can be found by differentiation with respect to the appropriate source.
These have the form
\begin{equation}
\langle \ol \psi_a \psi_a \rangle
=
\frac{1}{4} 
\tr 
\left[ K(J_a, m_a)^{-1} 
\begin{pmatrix}
0 & 1 \\
-1 & 0
\end{pmatrix}
\right], 
\text{ and }
\langle S_{aa} \rangle
=
\frac{1}{2}
\tr
\left[
K(J_a, m_a)^{-1}
\begin{pmatrix}
C \gamma_5 & 0 \\
0 & 0 
\end{pmatrix}
\right]
\label{eq:KH}
.\end{equation}
At zero quark mass, 
and zero diquark source, 
the chiral condensate is proportional to the density of zero modes of the operator 
$\cD (0)$%
~\cite{Banks:1979yr}. 
In this limit, 
the classic result is recovered from Eq.~\eqref{eq:KH}. 
Analogously, 
the diquark condensate can be non-vanishing as
$J_a \to 0$
if the operator
$K$ 
has zero modes. 
We need not be rigorous here, 
because we do not attempt to prove diquarks condense, 
or that chiral symmetry is spontaneously broken. 
We take note of these possibilities and handle them accordingly.

Now consider the difference of diquark condensates among two flavors. 
We wish to show that the difference vanishes in the limit of vanishing diquark sources
$J_a = J_b = 0$,
and degenerate masses. 
For a fixed gauge and auxiliary field background, 
this should be the case, 
because we can formally write the difference in the form
\begin{eqnarray}
\langle S_{aa} \rangle 
- 
\langle S_{bb} \rangle 
&=&
\frac{1}{2} 
\tr
\Bigg\{
\Big(
\left[
K(J_a, m)^{-1}
-
K(0,m)^{-1}
\right]
\notag \\
&& \phantom{space}
-
\left.
\left[ 
K(J_b,m)^{-1}
- 
K(0,m)^{-1}
\right]
\Big)
\begin{pmatrix}
C \gamma_5 & 0 \\
0 & 0
\end{pmatrix}
\right\}
.\end{eqnarray}
Each of the bracketed terms has no zero mode, 
so there are no singularities to worry about as
$J_a,  J_b \to 0$.  
Hence we have
$\langle S_{aa} \rangle - \langle S_{bb} \rangle = 0$.

We must also verify that there is no flavor breaking for the chiral condensate in the presence of diquark sources. 
The difference of chiral condensates among two flavors can be formally written as
\begin{eqnarray}
\langle \ol \psi_a \psi_a \rangle 
-
\langle \ol \psi_b \psi_b \rangle
&=&
\frac{1}{4} 
\tr
\Bigg\{
\Big(
\left[
K(J_a, m_a)^{-1}
- 
K(0, m_a)^{-1}
\right]
-
\left[
K(J_b, m_b)^{-1}
-
K(0, m_b)^{-1}
\right]
\notag \\
&& \phantom{space}
+
\left[
K(0,m_a)^{-1} 
- 
K(0, m_b)^{-1}
\right]
\left.
\Big)
\begin{pmatrix}
0 & 1 \\
-1 & 0 
\end{pmatrix}
\right\}
.\end{eqnarray}
Due to the subtractions, 
the first two bracketed terms have no zero modes, 
and the diquark sources can be safely taken to zero. 
The last bracketed term has no singularity as 
$m_b \to m_a$; 
it is precisely the case considered in~\cite{Vafa:1983tf}.
Hence we have
$\langle \ol \psi_a \psi_a \rangle - \langle \ol \psi_b \psi_b \rangle = 0$.

At this point, 
all we have assumed is that potentially divergent contributions to condensate differences can be regulated in a straightforward manner, which is certainly highly plausible. Assuming that this regularization has been done, to  show that the resulting differences of bpion condensates and chiral condensates vanish, 
we must integrate over the gauge field as well as the auxiliary fields weighted by the exponential of their action, 
and a factor of
$\Pf [ K (J_a, m_a) ]$
for each flavor. 
Here there is the additional complication that we are not always guaranteed a positive integration measure from the Pfaffian. 
In the case of the undeformed theory, 
one can show the Pfaffian is positive, and the equality of condensates should survive averaging over the gauge fields.
For the deformed theories, where the Pfaffian is not always positive,
it seems plausible that the fixed background result should survive integration over the gauge and auxiliary fields, 
but this is hard to prove.

To summarize:  if we assume non-vanishing chiral and diquark condensates, 
we have argued that they are plausibly flavor blind when the masses are degenerate, 
i.e.~%
$\langle \ol \psi_a \psi_b \rangle \propto \d_{ab}$, 
and 
$\langle S_{ab} \rangle \propto \d_{ab}$.%
\footnote{
The off-diagonal terms of the chiral condensate are obviously zero.
The off-diagonal diquark condensates also vanish,  
which can be easily demonstrated by performing the contractions in the Nambu-Gor'kov representation. 
For this reason, 
we have only considered diagonal diquark sources in Eq.~\eqref{eq:sourced}. 
} 
While the former is invariant under 
$U( N_f)_V$
transformations, 
the latter is only invariant under
$SO( N_f)_V$
transformations. 
Thus our argument amounts to the conjecture that 
$SO( N_f)_V$
symmetries cannot be spontaneously broken. 
We encountered two illustrations of this conjecture above in Sec.~\ref{sec:VacuumOrientation}
where we considered vacuum alignment in the presence of chirally symmetric and non-symmetric deformations. 
For the chirally non-symmetric deformation, 
the argument becomes more rigorous provided one is close enough to the chiral limit to ensure
a positive integration measure. 
The considerations in main text, 
however, 
were made without recourse to the arguments in this appendix.  
Instead, in Sec.~\ref{sec:VacuumOrientation}
we analyzed the vacuum alignment using the effective theory, 
and proved the non-breaking of 
$SO(N_f)_V$ 
for both deformations, including away from the chiral limit. 
Away from the chiral limit,
it is not known how auxiliary fields can be introduced for either the $V_{+}$ or the $V_{-}$ deformations while maintaining a positive integration measure. 
Nonetheless, 
the conjecture holds in the EFT.

\section{Vacuum alignment for the chirally non-symmetric deformation} %
\label{AppendixB} 		       				                        %

In Sec.~\ref{sec:ChiralOdd}, 
we gave the vacuum alignment for the case of the chirally non-symmetric deformation. 
Here we provide the technical details concerning the vacuum minimization. 
In Eq.~\eqref{eq:VminusEnergyDensity2}, 
the scaled action density for the vacuum, 
$s_-(a, \varphi)$, 
is reduced to a function of just 
two parameters:
$a$ 
and 
$\varphi$.
The parameter 
$a$ 
is bounded by unitarity. 
First we consider the endpoint, 
$a = 1$, 
subject to 
$\frac{\partial s_- }{ \partial \varphi} = 0$, 
and
$\frac{\partial^2 s_-}{\partial \varphi^2} > 0$.  
This singles out only the value
$\varphi = 0$, 
for which the value of the action is
$s_-(1, 0) = - 1$. 
The endpoint hence corresponds to the phase for which the orbifold equivalence holds.

In the interior, 
the critical points are found by solving the simultaneous equations:
$\frac{\partial s_-}{ \partial a} = \frac{\partial s_- }{ \partial \varphi} = 0$.  
For generic values of the low-energy parameters, 
there is only one solution with 
$a > 0$,
namely
$\varphi = 0$,
and
$a = a_-$, 
with 
$a_-$
given in Eq.~\eqref{eq:AMinus}. 
At this critical point, 
the mixed second derivative vanishes, 
$\frac{\partial^2 s_-}{\partial a \partial \varphi} = 0$.  
Requiring 
$\frac{\partial^2 s_-}{\partial a^2} > 0$, 
forces 
$a_- > 0$
for the point 
$\varphi = 0$, 
$a = a_-$ 
to be a minimum. 
We must also have
$\frac{\partial^2 s_-}{\partial \varphi^2} > 0$
to rule out a saddle point. 
This will automatically be satisfied for 
$0 < a_- < 1$, 
where the upper bound 
on 
$a_-$
follows from unitarity.   
The value of the vacuum energy in this phase is
$s_-(a_-, 0) = - \frac{1}{2} ( a_- + a_-^{-1}) < - 1$. 
Bpions are condensed in this phase.

For very special values of the low-energy parameters, 
there is an additional local minimum of the action.   
When 
$\varphi \neq 0$
and 
$\varphi \neq \pi$, 
one can satisfy the simultaneous equations
\begin{align}
\frac{\partial s_- }{ \partial a} 
&=
a ( x - y \cos 2 \varphi) - \cos \varphi
=
0,
\notag \\
\frac{\partial s_- }{ \partial \varphi} 
&=
\left[
a - 2 y ( 1 - a^2) \cos \varphi 
\right]
\sin \varphi
= 0
\label{eq:simult}
,\end{align}
with the values
$a_c$ 
and 
$\varphi_c$
given by
\begin{align} \label{eq:Ac}
a_c
&=
\left[ 1 - \frac{1}{\sqrt{2 y ( x + y)}} \right]^{\frac{1}{2}},
\\
\cos
\varphi_c \label{eq:Phic}
&=
a_{c} \,
\sqrt{\frac{x+y}{2 y}}
.\end{align}
While there are four possible solutions for 
$a_c$
in Eq.~\eqref{eq:simult},
only the one shown in Eq.~\eqref{eq:Ac}
can satisfy
$0 < a_c < 1$. 
This occurs when
$2 y ( x + y) > 1$, and in fact the curve $2y(x+y)=1$ is the lower boundary of the metastable phase seen in Figure~\ref{fig:PhaseDiagram}. 
The low-energy parameters 
$x$, 
and
$y$
must also be such that
$\cos \varphi_c  < 1$, 
and provided this condition is met, Eq.~\eqref{eq:Ac} determines the angle 
$\varphi_c$
up to sign.

The restrictions 
$0 < a_c < 1$, 
and
$0 < \cos \varphi_c < 1$
are already enough to show that both
$\frac{\partial^2 s_-}{\partial a^2}  >0$
and
$\frac{\partial^2 s_-}{\partial \varphi^2}  >0$.
In order for the point
$(a_c, \varphi_c)$
to be a minimum, 
it must be the case that
the discriminant $\mathfrak{D}$ satisfies
$\mathfrak{D} >0$, 
where
\begin{align}
\mathfrak{D}
&\equiv
\frac{\partial^2 s_-}{\partial a^2} \frac{\partial^2 s_-}{\partial \varphi^2} 
- 
\frac{\partial^2 s_-}{\partial a \partial \varphi}
,\end{align}
to exclude saddle points.
Quite tediously, 
one can show that positivity of the discriminant forces 
$\varphi_c < 0$. 
(It is obvious that 
$\mathfrak{D} > 0$
when 
$\varphi_c < 0$; 
the tedium arises in showing this condition is necessary.)

When it exists, 
this exotic parity and $U(1)_{B}$-breaking phase is 
\emph{always}
metastable. 
To investigate this, 
we note that the energy density takes a simple form at the local minimum
\begin{equation}
\label{eq:ExoticPhaseEnergy}
s_-( a_c, \varphi_c)
=
- \sqrt{\frac{x+y}{2 y}} + \frac{1}{4 y} 
.\end{equation}
When
$y > x - 1$, 
the energy of the exotic phase must compete with that in the normal phase,
$s_-(1,0) = -1$.  The energy surfaces $s_-(1,0)$ and $s_-(a_c, \varphi_c)$ intersect along the two curves defined by
$x = y + 1 + \frac{1}{8y}$, but it is easy to see that both of these curves lie outside of the region  defined by $y> x - 1$ and $x > 0$. Therefore there is no crossing of phases to worry about in this regime. Evaluating the normalized energies at the point
$(x,y) = (1,1)$, we see the normal phase wins. 
Since there is no crossing of phases, this is enough to prove that the normal phase wins wherever it coexists with the exotic phase.  

On the other hand, 
when 
$y < x - 1$, 
the exotic phase competes with the bpion condensed phase, 
which has the energy density 
\begin{equation}
s_-(a_-, 0) 
= - \frac{1}{2} \left( \frac{1}{x-y} + x - y \right)
.\end{equation} 
To see that the exotic phase always loses to the bpion-condensed phase, note that the surfaces defined by $s_{-}(a_{-},0)$ and $s_-( a_c, \varphi_c)$ intersect along two curves $l_{\pm}$ defined by
\begin{align}
l_{\pm}:  x_{\pm} = \frac{4 y^{2} +1  \pm \sqrt{16 y^{2}+1}}{4y} .
\end{align}
It is not hard to show that $l_{-}$ lies in the region in the $(x,y)$ plane defined by $y> x-1$, so it is not relevant in our current discussion.  However, one can verify through some rather tedious algebra that $\cos \varphi_{c} = 1$ along the curve $l_{+}$.   For all points $(x,y)$ which lie to the right this curve, $\cos \varphi_{c} > 1$ and the metastable phase does not exist.  To the left of $l_{+}$, we have $\cos{\varphi_{c}} < 1$, and until one reaches the curve $2y(x+y)=1$, it is also the case that $a_{c}<1$, so that these two curves define the boundaries of the metastable phase.  We can verify that the exotic phase is indeed metastable everywhere in the bpion-condensed region by comparing the values of $s_{-}(a_{-},0)$ and $s_-( a_c, \varphi_c)$ near  $l_{+}$, at $(x,y) =(x_{+}+\epsilon,y)$:
\begin{align}
s_{-}(a_{-},0)&= -\frac{f(y)}{4y} -\frac{\epsilon}{1+ f(y)} - \frac{32 y^{3} \epsilon^{2}}{[1+f(y)]^{3}}+  \mathcal{O}(\epsilon^{3}) \\
s_{-}(a_{c},\varphi_{c})&= -\frac{f(y)}{4y} -\frac{\epsilon}{1+ f(y)} + \frac{y \epsilon^{2}}{\sqrt{2}\left[\frac{1}{2}f(y)^{2} +\frac{1}{2} +f(y)\right]^{3/2}} + \mathcal{O}(\epsilon^{3}),
\end{align}	
where $f(y) = \sqrt{16 y^{2}+1}$.  Clearly the bpion-condensed phase has a lower energy than the exotic phase near $l_{+}$.  Together with the information on where the energy surfaces of the bpion-condensed and the exotic phase intersect, this is enough to conclude that the bpion-condensed always remains the ground state when 
$y < x - 1$.  As advertised the exotic phase is metastable everywhere it exists.

\bibliography{orbifoldingNoArxiv}

\end{document}